\documentclass[final, authoryear, 5p, times]{elsarticle}
\newcommand{\sbtitle}{A variance reduction strategy for numerical random homogenization based on the equivalent inclusion method}
\newcommand{\sbjournal}{unspecified}

\usepackage[fleqn]{amsmath}

\usepackage{amssymb}
\usepackage{amsthm}
\usepackage{bm}
\usepackage[breaklinks=true,colorlinks=true,linktocpage=true,pdftitle={\sbtitle}, pdfauthor={S. Brisard, M. Bertin and F. Legoll},urlcolor=blue]{hyperref}

\usepackage{pgf}

\usepackage{textcomp}

\newcommand{\app}{\text{app}}

\DeclareMathOperator{\besselj}{J}

\DeclareMathOperator{\corr}{corr}
\DeclareMathOperator{\cov}{cov}

\newcommand{\D}{\mathrm d}

\renewcommand{\div}{\operatorname{div}}
\newcommand{\E}{\mathrm e}
\newcommand{\eff}{\mathrm{eff}}
\newcommand{\EIM}{\mathrm{EIM}}
\DeclareMathOperator{\ensavg}{\mathbb E}
\newcommand{\FEM}{\mathrm{FEM}}

\DeclareMathOperator{\grad}{\vec{grad}}

\newcommand{\I}{\mathrm i}
\newcommand{\integers}{\mathbb Z}

\newcommand{\per}{\mathrm{per}}

\newcommand{\reals}{\mathbb R}

\newcommand{\tens}[2][]{\vec{#2}}

\DeclareMathOperator{\tr}{tr}

\DeclareMathOperator{\var}{var}
\renewcommand{\vec}[1]{\bm{\mathrm{#1}}}

\newcommand{\cell}{\mathcal Q}

\newcommand{\eps}{\varepsilon}

\newtheorem{remark}{Remark}
\newtheorem{problem}{Problem}

\usepackage{xcolor}
\definecolor{PodoOrange}{RGB}{230, 159, 0}
\definecolor{PodoSkyBlue}{RGB}{86, 180, 233}
\definecolor{PodoBluishGreen}{RGB}{0, 158, 115}
\definecolor{PodoYellow}{RGB}{240, 228, 66}
\definecolor{PodoBlue}{RGB}{0, 114, 178}
\definecolor{PodoVermillion}{RGB}{213, 94, 0}
\definecolor{PodoReddishPurple}{RGB}{204, 121, 167}

\definecolor{sborange}{named}{PodoOrange}
\definecolor{sblightblue}{named}{PodoSkyBlue}
\definecolor{sbgreen}{named}{PodoBluishGreen}
\definecolor{sbyellow}{named}{PodoYellow}
\definecolor{sbblue}{named}{PodoBlue}
\definecolor{sbred}{named}{PodoVermillion}
\definecolor{sbpurple}{named}{PodoReddishPurple}

\definecolor{vlightgray}{RGB}{240, 240, 240}

\journal{\sbjournal}
\newlength{\pgfcadvthin}
\newlength{\pgfcadthin}
\newlength{\pgfcadnormal}
\newlength{\pgfcadthick}
\newlength{\pgfcadvthick}
\newlength{\pgfcadrightanglemarksize}
\newcommand{\pgfcadmarkangle}[4]{
  \begin{pgfscope}
    \pgfsetarrowsend{stealth}
    \pgfpathmoveto{\pgfpoint{#2}{0}}
    \pgfpatharc{0}{#1}{#2}
    \pgfusepath{stroke}
    \begin{pgfscope}
      \pgftransformarcaxesattime{0.5}{\pgfpointorigin}{\pgfpoint{#2}{0}}{\pgfpoint{0}{#2}}{0}{#1}
      \pgfcoordinate{pgfxmarkanglep1}{\pgfpointorigin}
    \end{pgfscope}
    \pgftransformarrow{pgfpointorigin}{\pgfpointanchor{pgfxmarkanglep1}{center}}
    \pgftransformrotate{-180}
    \pgfnode{rectangle}{#3}{#4}{}{\pgfusepath{}}
  \end{pgfscope}
}

\newlength{\pgfcaddimlegsize}
\newcommand{\pgfcaddim}[5]{
  \begin{pgfscope}
    \pgftransformlineattime{0}{#1}{#2}
    \pgftransformshift{\pgfpoint{0}{#3}}
    \pgfcoordinate{pgfxp1}{\pgfpointorigin}
    \pgfpathmoveto{\pgfpoint{0}{-\pgfcaddimlegsize}}
    \pgfpathlineto{\pgfpoint{0}{\pgfcaddimlegsize}}
    \pgfpathmoveto{\pgfpoint{-\pgfcaddimlegsize}{0}}
    \pgfpathlineto{\pgfpointorigin}
    \pgfusepath{stroke}
  \end{pgfscope}
  \begin{pgfscope}
    \pgftransformlineattime{1}{#1}{#2}
    \pgftransformshift{\pgfpoint{0}{#3}}
    \pgfcoordinate{pgfxp2}{\pgfpointorigin}
    \pgfpathmoveto{\pgfpoint{0}{-\pgfcaddimlegsize}}
    \pgfpathlineto{\pgfpoint{0}{\pgfcaddimlegsize}}
    \pgfpathmoveto{\pgfpointorigin}
    \pgfpathlineto{\pgfpoint{\pgfcaddimlegsize}{0}}
    \pgfusepath{stroke}
  \end{pgfscope}
  \begin{pgfscope}
    \pgfsetarrows{stealth-stealth}
    \pgfpathmoveto{\pgfpointanchor{pgfxp1}{center}}
    \pgfpathlineto{\pgfpointanchor{pgfxp2}{center}}
    \pgfusepath{stroke}
    \pgftransformlineattime{0.5}{\pgfpointanchor{pgfxp1}{center}}{\pgfpointanchor{pgfxp2}{center}}
    \pgfnode{rectangle}{#4}{#5}{}{\pgfusepath{}}
  \end{pgfscope}
}

\newcommand{\pgfcaddecoratedline}[5]{
  \begin{pgfscope}
    \pgfpathmoveto{#2}
    \pgfpathlineto{#3}
    \pgfusepath{stroke}
    \pgftransformlineattime{#1}{#2}{#3}
    \pgfnode{rectangle}{#4}{#5}{}{\pgfusepath{}}
  \end{pgfscope}
}

\newlength{\sbfigwidth}
\newlength{\sbL}
\newlength{\sbunitvectorlength}
\newlength{\sbdimoffset}
\begin{document}

\begin{frontmatter}

  \title{\sbtitle}
  \author[adr:1]{S\'ebastien Brisard\corref{cor:1}}\ead{sebastien.brisard@univ-eiffel.fr}
  \author[adr:1,adr:2,adr:3]{Micha\"el Bertin}
  \author[adr:1,adr:3]{Fr\'ed\'eric Legoll}\ead{frederic.legoll@enpc.fr}
  \address[adr:1]{Navier, Ecole des Ponts, Univ Gustave Eiffel, CNRS, Marne-la-Vall\'ee, France}
  \address[adr:2]{CERMICS, Ecole des Ponts, Marne-la-Vall\'ee, France}
  \address[adr:3]{MATHERIALS project-team, Inria, Paris, France}
  \cortext[cor:1]{Corresponding author.}

  \begin{abstract}
    Using the equivalent inclusion method (a method strongly related to the Hashin-Shtrikman variational principle) as a surrogate model, we propose a variance reduction strategy for the numerical homogenization of random composites made of "inclusions" (or rather inhomogeneities) embedded in a homogeneous matrix. The efficiency of this strategy is demonstrated within the framework of two-dimensional, linear conductivity. Significant computational gains vs full-field simulations are obtained even for high contrast values. We also show that our strategy allows to investigate the influence of parameters of the microstructure on the macroscopic response. Our strategy readily extends to three-dimensional problems and to linear elasticity. Attention is paid to the computational cost of the surrogate model. In particular, an inexpensive approximation of the so-called influence tensors (that are used to compute the surrogate model) is proposed.
  \end{abstract}

  \begin{keyword}
    Homogenization \sep
    Full-field simulations \sep
    Conductivity \sep
    Variance reduction \sep
    Surrogate model
  \end{keyword}

\end{frontmatter}

\section{Introduction}

Most natural and manufactured materials are heterogeneous, and the prediction of their effective properties must account for this small-scale heterogeneity by means of an appropriate upscaling strategy. When the classical mean-field/effective-field techniques (e.g. the Hashin-Shtrikman bounds, the Mori-Tanaka bounds, the Self-Consistent scheme), fail to deliver a prediction that is accurate enough, expensive full-field numerical simulations must be performed. These simulations compute the solution to the so-called \emph{corrector problem} within an ``elementary cell'' (to be defined below), under specific local constitutive laws. The elementary cell is subjected to an elementary macroscopic loading that ensures that the Hill--Mandel lemma holds.

The extent of the elementary cell depends on the geometric structure of the heterogeneities. For \emph{periodic homogenization}, the elementary cell is well-defined and corresponds to the unit-cell~\citep[Chap.~1]{Allaire02}, \citep{blp78,milt2002,sanchez80,sanchez92}. For \emph{random homogenization} (the topic of the present article), the elementary cell should be an infinite realization of the random material~\citep{jiko1994,milt2002}, which is of course not practical. The computation is therefore carried out on a large, but finite-size realization of the random material, which we call here a \emph{statistical volume element} (SVE), following~\citet{osto2006}. The solution to the corrector problem on the SVE delivers the \emph{apparent properties}. In the limit of asymptotically large SVEs, the apparent properties converge to the true \emph{effective (homogenized) properties}.

The apparent properties thus defined appear as a size-depen\-dent random variable. The above-discussed workflow introduces two types of errors: \textit{(i)}~a deterministic bias (that relates to the finite size of the SVE: the average of the apparent properties over all possible microstructures in the SVE is {\rm different} from the exact effective, homogenized properties) and \textit{(ii)}~random errors (that relate to the realization itself: the microstructure in the SVE being random, its associated apparent properties are random). 
In this article, because this is overwhelmingly the case in the numerical practice, we consider corrector problems in the SVE complemented by {\em periodic boundary conditions}. This choice is motivated by considerations about the decrease of the systematic error (i.e. the deterministic bias) when the SVE size increases. It has been numerically observed (see e.g.~\citet{kani2003}) and theoretically demonstrated in some cases (see~\cite{nolen14} and the works of A.~Gloria, F.~Otto and their collaborators initiated in~\citet{gloria11}) that the statistical error decays with a slower rate (with respect to the SVE size) than the systematic error. For large SVEs, the statistical error is therefore dominating the systematic error. In what follows, we focus on reducing the statistical error, and describe an approach to more efficiently compute the \emph{average apparent properties}.

\medskip

Estimating the average apparent properties naturally requires the Monte--Carlo method, which is known to converge slowly. Indeed, the central limit theorem delivers an estimate of the confidence interval that scales as \(M^{-1/2}\), where \(M\) is the number of realizations (the number of full-field simulations in the present case). In other words, dividing the amplitude of the confidence interval by 10 requires to multiply the number of realizations by 100. This can lead to costly simulations.


Observing that the amplitude of the confidence interval is also proportional to the standard deviation of the random variable under consideration, variance reduction techniques aim at refining the estimate of the statistical error by controlling this standard deviation rather than the sample size. Many types of such approaches have been proposed in the literature (antithetic variables, control variate, importance sampling, stratification based approaches, to name but a few), in various contexts where the Monte--Carlo method is used (we refer to~\citet{fishman1996} for a review of these approaches). We mention here that we have already adapted some of these methods to the present context of numerical random homogenization, for linear and for some nonlinear problems, in~\citet{costaouec_mprf,costaouec_cedya,costaouec_lncse,sqs2016,lego2015,dcds2015} (see also the review articles~\citet{blan2016,lebris-legoll2017}).

In this article, we focus on the class of control variate approaches, which are based on the use of a so-called \emph{surrogate model}. This new random variable should be strongly correlated to the random variable of interest and its expectation should be known exactly (alternatively, evaluation of the surrogate model should be inexpensive enough to allow for a very accurate Monte--Carlo estimation of its expectation). The Monte--Carlo method is then run on a linear combination of the random variable of interest with the surrogate model, the coefficients of the linear combination being chosen such that its standard deviation is smaller than that of the random variable of interest.

Within the framework of linear electric conductivity, we present here a control-variate-based variance reduction strategy for the determination of the effective conductivity of an assembly of circular inclusions embedded in a homogeneous matrix. We do not consider size-effects, which (as argued previously) are a separate issue. We therefore work on \emph{fixed-size} SVEs and focus on the determination of the \emph{average apparent conductivity}.

We use the equivalent inclusion method (EIM) to produce inexpensive estimates of the apparent conductivity. The EIM was first introduced within the framework of linear elasticity by~\citet{mosc1975}. It relies on the formulation of the corrector problem as an integral equation known as the Lippmann--Schwinger equation~\citep{korr1973,zell1973,kron1974}. As suggested by~\citet{mosc1975}, the EIM seeks an approximate solution to the Lippmann--Schwinger equation that is constant in each inclusion. Since their seminal work, various strategies (mostly relying on collocation) have been proposed to derive EIM equations. Following~\citet{bris2014}, we use here a Galerkin approach in a periodic setting.

The EIM estimates of the apparent conductivity then serve as a surrogate model which is used in conjunction with full-field simulations performed using the finite element method (FEM). We show that this surrogate model is indeed effective at reducing the amplitude of the confidence interval, even in the difficult cases of very large contrast between the matrix and the inclusions. In order for the surrogate model to also be cost-effective, an inexpensive approximation of the so-called influence tensors is proposed. This approximation sheds new light on the recent work by~\citet{zece2021}, where a similar approximation is considered.

We underline here the strong mechanical content of the EIM, which is a surrogate model well-tailored to our application, namely the homogenization of random assemblies of inclusions embedded in a homogeneous medium. We also emphasize the strong link between EIM, the Hashin-Shtrikman variational principle and the Hashin-Shtrikman bounds. This link is further discussed in Sec.~\ref{sec:eim_method}.

\medskip

The article is organized as follows. Variance reduction techniques based on the use of a surrogate model (a.k.a. control variate techniques) are briefly reviewed in Sec.~\ref{sec:20201013092415}. Then, Section~\ref{sec:20220801153432} provides an overview of the equivalent inclusion method, in particular its variational form. Our variance reduction strategy is outlined in Sec.~\ref{sec:20220801101220} where some specificities of the EIM are underlined. The efficiency of the proposed strategy is demonstrated by numerical experiments in a two-dimensional setting in Sec.~\ref{sec:20220727145908}. We collect some conclusions in Sec.~\ref{sec:conc}.

\bigskip

Note that all data necessary to reproduce the results presented in Sec.~\ref{sec:20220727145908} and Appendix~\ref{sec:20220719172106} are available at the \texttt{Recherche Data Gouv} open data repository~\citep{bris2023, bris2023a}.

\section{Variance reduction through a surrogate model} \label{sec:20201013092415}

In the present section, we give a brief overview of the control variate variance reduction technique. Our goal is to compute a numerical estimate of the ensemble average \(\ensavg(X)\) of a scalar random variable \(X(\omega)\). To do so, the Monte--Carlo approach consists in drawing \(M\) independent realizations of $X(\omega)$, denoted \(X_1(\omega)\), \ldots, \(X_M(\omega)\), and evaluating the empirical average
$$
\mu_M(X) = \frac{1}{M} \sum_{m=1}^M X_m(\omega).
$$
The law of large numbers says that the empirical average $\mu_M(X)$ converges, when $M \to +\infty$, to the exact ensemble average \(\ensavg(X)\). The empirical average $\mu_M(X)$ is unbiased, in the sense that, for any $M$, its expectation is equal to \(\ensavg(X)\). The only error between $\mu_M(X)$ and \(\ensavg(X)\) is therefore a statistical error, which is estimated by the central limit theorem:
$$
\lim_{M\to+\infty} \Pr\left[ \frac{1}{M} \sum_{m=1}^M X_m(\omega) - \ensavg(X) \leq \alpha \, \frac{\sigma_X}{M^{1/2}} \right] = \Phi(\alpha),
$$
where \(\Phi\) is the standard normal cumulative distribution function and \(\sigma_X^2\) is the variance of \(X\) (\(\sigma_X\) is the so-called standard deviation of \(X\)). The statistical error thus scales as \(\alpha \, \sigma_X \, M^{-1/2}\). For example, for the \(99\,\%\) confidence interval, \(\alpha=2.6\): with a probability of \(99\,\%\), the expectation $\ensavg(X)$ belongs to the interval
$$
\left[ \frac{1}{M} \sum_{m=1}^M X_m(\omega) - 2.6 \frac{\sigma_X}{M^{1/2}},\frac{1}{M} \sum_{m=1}^M X_m(\omega) + 2.6 \frac{\sigma_X}{M^{1/2}} \right].
$$
As expected, increasing the number \(M\) of samples reduces the statistical error, albeit at the slow rate \(M^{-1/2}\).

As pointed out above, several variance reduction strategies have been proposed in the literature with the aim to provide a more efficient approach than the Monte--Carlo method described above. The control variate approach \citep[see e.g.][p.~277]{fishman1996} is one of them. It is based on the observation that a reduction of the statistical error can be obtained by operating on the variance of \(X\) itself. More precisely, let us consider a second random variable \(Y(\omega)\), denoted below the surrogate model. A crucial assumption is that the ensemble average \(\ensavg(Y)\) is \emph{known}. We then consider the new random variable
$$
Z(\omega) = X(\omega) - \xi \, \bigl[ Y(\omega)-\ensavg(Y) \bigr],
$$
where \(\xi\) is a free (deterministic) parameter. Since \(Z\) and \(X\) have the same ensemble average, we are left with evaluating the ensemble average of \(Z\). To do so, we again compute an empirical average
$$
\mu_M(Z) = \frac{1}{M} \sum_{m=1}^M Z_m(\omega),
$$
where the realizations $Z_m(\omega)$ are given by $Z_m(\omega) = X_m(\omega) - \xi \, \bigl[ Y_m(\omega)-\ensavg(Y) \bigr]$. Using that $\ensavg(Z) = \ensavg(X)$, the central limit theorem yields
$$
\lim_{M\to+\infty} \Pr\left[ \frac{1}{M} \sum_{m=1}^M Z_m(\omega) - \ensavg(X) \leq \alpha \, \frac{\sigma_Z}{M^{1/2}} \right] = \Phi(\alpha).
$$
Note that the accuracy is now controlled by the standard deviation $\sigma_Z$ of the random variable $Z$. The variance of $Z$ is given by
$$
\var(Z) = \var(X) - 2 \, \xi \cov(X,Y) + \xi^2 \var(Y),
$$
where \(\cov(X,Y) = \ensavg \bigl[ (X-\ensavg(X)) \, (Y-\ensavg(Y)) \bigr]\) denotes the covariance of \(X\) and \(Y\). Therefore, any choice of \(\xi\) that makes \(\xi^2 \var(Y) - 2 \, \xi \, \cov(X,Y) \leq 0\) effectively reduces the statistical error on the numerical estimation of \(\ensavg(X)\). The optimal choice \(\xi^\star\) of \(\xi\) is the one which minimizes \(\var(Z)\):
\begin{equation} \label{eq:def_xi_star}
\xi^\star = \frac{\cov(X,Y)}{\var(Y)}.
\end{equation}
The optimal variance of \(Z\) reads
\begin{equation*} 
  \var(Z^\star) = \bigl[1-\corr(X,Y)^2\bigr] \var(X),
\end{equation*}
where \(\corr(X,Y) = \cov(X,Y) \, \bigl[\var(X)\var(Y)\bigr]^{-1/2}\) denotes the correlation of \(X\) and \(Y\) (by construction, $| \corr(X,Y) | \leq 1$). The above equation shows that, if \(X\) and \(Y\) are strongly correlated, then the variance of \(Z^\star\) (hence the statistical error on the numerical estimate of its ensemble average) is much smaller than the variance of \(X\).

In order to evaluate the ensemble average of \(X\), the Monte--Carlo strategy is therefore applied to \(Z\) (with reduced statistical error). The quantity of interest \(\ensavg(X)\) is then obtained from the identity \(\ensavg(X) = \ensavg(Z)\).

In practice, the quantities \(\cov(X,Y)\) and \(\var(Y)\), which are needed to find the optimal value of $\xi$ (see~\eqref{eq:def_xi_star}), are not known. They are estimated numerically as follows:
$$
\cov(X,Y) \simeq \frac{1}{M} \sum_{m=1}^M X_m(\omega) Y_m(\omega) - \frac{1}{M^2} \Bigl( \sum_{m=1}^M X_m(\omega) \Bigr) \Bigl( \sum_{m=1}^M Y_m(\omega) \Bigr),
$$
and
$$
\var(Y) \simeq \frac{1}{M} \sum_{m=1}^M (Y_m(\omega))^2 - \frac{1}{M^2} \left( \sum_{m=1}^M Y_m(\omega) \right)^2.
$$
In practice, the random variable $Y$ is often cheap to evaluate in comparison to the evaluation of $X$. The cost to compute the empirical average $\frac{1}{M} \sum_{m=1}^M X_m(\omega)$ is therefore identical to the cost to compute $\frac{1}{M} \sum_{m=1}^M Z_m(\omega)$. The latter estimate is a more accurate approximation of $\ensavg(X)$ than the former: after variance reduction, and at equal computational cost, the statistical error on \(\ensavg(X)\) is indeed reduced by the factor
$$
\frac{\sigma_X}{\sigma_{Z^\star}} = \sqrt{\frac{1}{1-\corr(X,Y)^2}},
$$
defined as the \emph{efficiency index} (higher is better).

\begin{remark} \label{rem:expect_unknown}
In the case when the expectation of $Y$ is not known, it is still possible to use the above control variate approach by replacing $\ensavg(Y)$ by an accurate estimate of the form $\displaystyle \frac{1}{\cal M} \sum_{m=1}^{\cal M} Y_m(\omega)$, where the number ${\cal M}$ of realizations that are considered is sufficiently large such that the additional statistical error introduced by this approximation is much smaller than the statistical error between $\mu_M(X)$ and $\ensavg(X)$. We typically take ${\cal M} \gg M$. In the case when $Y$ is inexpensive to evaluate, the additional cost is negligible.
\end{remark}

\section{Overview of the equivalent inclusion method for homogenization} \label{sec:20220801153432}

In this section, we first present the electric conductivity problem of interest in this article, as well as the associated homogenization results (see Sec.~\ref{sec:20200514070529}). We next turn to the equivalent inclusion method (in Sec.~\ref{sec:20201006103812}), which is a strategy to approximate the apparent conductivity that we are going to use here as a surrogate model, as explained in Sec.~\ref{sec:20220801101220}.

\subsection{Effective and apparent conductivity} \label{sec:20200514070529}

The homogenization problems considered in the present article are expressed within the framework of electric conductivity. We recall here this setting as well as the notion of effective and apparent conductivities.

We consider a random, heterogeneous material with local conductivity \(\tens{\sigma}_\eps(\vec x, \omega)\) (second-order, symmetric, positive definite tensor) where \(\vec x \in \reals^d\) denotes the observation point, while \(\omega\) refers to the realization. The subscript $\eps$ highlights the fact that the characteristic scale $\eps$ of variations of the conductivity is small. We consider the classical random ergodic homogenization setting, in which $\tens{\sigma}_\eps(\vec x, \omega) = \tens{\sigma}(\vec x/\eps, \omega)$ for a fixed stationary tensor $\tens{\sigma}$, that we assume to be (uniformy in $\vec x$ and $\omega$) bounded from below and from above. Each realization of $\tens{\sigma}(\cdot, \omega)$ fills the whole space \(\reals^d\).

The heterogeneous electric conductivity problem is given by
$$
\div \vec J_\eps = f, \quad \vec J_\eps = -\tens{\sigma}_\eps \, \grad \Phi_\eps,
$$
in some computational domain ${\cal D}$, where $f$ is the load (which is assumed independent of $\eps$ and $\omega$), $\Phi_\eps$ is the electric potential, $-\grad \Phi_\eps$ is the electric field and $\vec J_\eps$ is the electric current. The above problem is complemented with some boundary conditions (say $\Phi_\eps(\vec x,\omega) = 0$ for any $\vec x \in \partial {\cal D}$). It is then well-known (see e.g. the textbooks~\citet{jiko1994,milt2002,livre_blanc_lebris}, the lectures~\citet[Part~7]{bris2022a} and also the review article~\citet{singapour} and the extensive bibliography contained therein) that the corresponding homogenized problem reads
\begin{equation} \label{eq:pb_homog}
\div \vec J^\eff = f, \quad \vec J^\eff = -\tens{\sigma}^\eff \, \grad \Phi^\eff,
\end{equation}
where $\tens{\sigma}^\eff$ is the effective (here constant and deterministic) electric conductivity. This effective conductivity is defined in terms of the so-called corrector functions $\Phi$, which are solutions to the following problem: for any fixed constant macroscopic electric field $\overline{\vec E} \in \reals^d$, find $\Phi$ such that
\begin{equation} \label{eq:pb_corr}
  \begin{array}{c}
    \div \vec J = 0, \quad \vec J = \tens{\sigma} \, \vec E, \quad \vec E = \overline{\vec E} - \grad \Phi \quad \text{in $\reals^d$}, \\
    \text{$\grad \Phi$ is stationary and $\ensavg(\grad \Phi) = 0$.}
  \end{array}
\end{equation}
In practice, solving~\eqref{eq:pb_corr} is not possible, since the problem is posed on the \emph{whole} space $\reals^d$. For practical computations, we therefore consider restrictions of this problem (and thus of the realizations of the conductivity $\tens{\sigma}$) to the finite-size domain \(\cell = (-L/2,L/2)^d\) for some large $L$. Following~\citet{osto2006}, we will call these restrictions \emph{statistical volume elements}.

Owing to the periodic boundary conditions to be applied here, it will be convenient to define \emph{homologous} points on opposite sides of the boundary \(\partial \cell\) as two points \(\vec x^\pm \in \partial \cell\) such that there exists \(i \in \{1, \ldots, d\}\) such that \(x_j^+ = x_j^- + \delta_{ij} \, L\) for all \(1 \leq j \leq d\) (see Fig.~\ref{fig:20220728164931}).

\begin{figure}[htbp]
  \centering
  \begin{pgfpicture}
    \pgfslopedattimetrue
    \pgfsetlinewidth{\pgfcadnormal}
    \begin{pgfscope}
      \pgfsetlinewidth{\pgfcadthick}
      \pgfsetstrokecolor{sblightblue}
      \pgfsetfillcolor{vlightgray}

      \pgfpathrectangle{\pgfpoint{-0.5\sbL}{-0.5\sbL}}{\pgfpoint{\sbL}{\sbL}}

      \pgfusepath{stroke, fill}
    \end{pgfscope}

    \begin{pgfscope}
      \pgfpathmoveto{\pgfpoint{-0.5\sbfigwidth}{0}}
      \pgfpathlineto{\pgfpointorigin}
      \pgfpathlineto{\pgfpoint{0}{-0.5\sbfigwidth}}

      \pgfpathmoveto{\pgfpoint{\sbunitvectorlength}{0}}
      \pgfpathlineto{\pgfpoint{0.5\sbfigwidth}{0}}

      \pgfpathmoveto{\pgfpoint{0}{\sbunitvectorlength}}
      \pgfpathlineto{\pgfpoint{0}{0.5\sbfigwidth}}

      \pgfusepath{stroke}

      \pgfsetstrokecolor{sbpurple}
      \pgfsetarrowsstart{stealth}
      \pgfsetarrowsend{stealth}
      \pgfpathmoveto{\pgfpoint{0.0cm}{\sbunitvectorlength}}
      \pgfpathlineto{\pgfpointorigin}
      \pgfpathlineto{\pgfpoint{\sbunitvectorlength}{0.0cm}}
      \pgfusepath{stroke}

      \begin{pgfscope}
        \pgftransformshift{\pgfpoint{0.5\sbunitvectorlength}{0}}
        \pgfnode{rectangle}{north}{\(\color{sbpurple}\vec e_1\)}{}{\pgfusepath{}}
      \end{pgfscope}
      \begin{pgfscope}
        \pgftransformshift{\pgfpoint{0}{0.5\sbunitvectorlength}}
        \pgfnode{rectangle}{east}{\(\color{sbpurple}\vec e_2\)}{}{\pgfusepath{}}
      \end{pgfscope}
    \end{pgfscope}

    \begin{pgfscope}
      \pgftransformshift{\pgfpoint{0}{0.3\sbL}}
      \begin{pgfscope}
        \pgftransformshift{\pgfpoint{-0.5\sbL}{0}}
        \pgfnode{rectangle}{center}{\(\bullet\)}{}{\pgfusepath{}}
        \pgfnode{rectangle}{east}{\(\vec x_1^-\)}{}{\pgfusepath{}}
      \end{pgfscope}
      \begin{pgfscope}
        \pgftransformshift{\pgfpoint{+0.5\sbL}{0}}
        \pgfnode{rectangle}{center}{\(\bullet\)}{}{\pgfusepath{}}
        \pgfnode{rectangle}{west}{\(\vec x_1^+\)}{}{\pgfusepath{}}
      \end{pgfscope}
    \end{pgfscope}

    \begin{pgfscope}
      \pgftransformshift{\pgfpoint{-0.3\sbL}{0}}
      \begin{pgfscope}
        \pgftransformshift{\pgfpoint{0}{-0.5\sbL}}
        \pgfnode{rectangle}{center}{\(\bullet\)}{}{\pgfusepath{}}
        \pgfnode{rectangle}{north}{\(\vec x_2^-\)}{}{\pgfusepath{}}
      \end{pgfscope}
      \begin{pgfscope}
        \pgftransformshift{\pgfpoint{0}{+0.5\sbL}}
        \pgfnode{rectangle}{center}{\(\bullet\)}{}{\pgfusepath{}}
        \pgfnode{rectangle}{south}{\(\vec x_2^+\)}{}{\pgfusepath{}}
      \end{pgfscope}
    \end{pgfscope}

    \begin{pgfscope}
      \pgfsetlinewidth{\pgfcadthin}
      \pgfcaddim{\pgfpoint{-0.5\sbL}{-0.5\sbL}}{\pgfpoint{0.5\sbL}{-0.5\sbL}}{-\sbdimoffset}{south}{\(L\)}
      \pgfcaddim{\pgfpoint{0.5\sbL}{0.5\sbL}}{\pgfpoint{0.5\sbL}{-0.5\sbL}}{\sbdimoffset}{south}{\(L\)}
    \end{pgfscope}

    \begin{pgfscope}
      \pgftransformshift{\pgfpoint{0.5\sbL}{-0.5\sbL}}
      \pgfnode{rectangle}{south east}{\(\cell\)}{}{\pgfusepath{}}
    \end{pgfscope}
  \end{pgfpicture}
  \caption{\((\vec x_1^-, \vec x_1^+)\) and \((\vec x_2^-, \vec x_2^+)\) are pairs of homologous points on the boundary of the two-dimensional domain $\cell$. \label{fig:20220728164931}}
\end{figure}

The \emph{apparent} conductivity \(\tens{\sigma}^\app(\omega,L)\) of the SVE is defined through the solution to the following boundary-value problem on \(\cell\), denoted as the corrector problem, the localization problem or the auxiliary problem in the literature:

\begin{problem}[Corrector problem] \label{pb:20201012143332}
  For any \emph{fixed} constant macroscopic electric field \(\overline{\vec E} \in \reals^d\), find the \(\cell\)-periodic fluctuations of the electric potential \(\Phi^\per\) such that
  \begin{gather}
    \label{eq:20200616062014}
    \div \vec J = 0 \quad \text{in $\cell$}, \\
    \label{eq:20200616062019}
    \vec J = \tens{\sigma} \, \vec E \quad \text{in $\cell$}, \\
    \label{eq:20200616062027}
    \vec E = \overline{\vec E} - \grad\Phi^\per \quad \text{in $\cell$}, \\
    \label{eq:20200514063224}
    \vec J^+ \cdot \vec n^+ = - \vec J^- \cdot \vec n^- \quad \text{on $\partial \cell$}.
  \end{gather}
  Equation~\eqref{eq:20200514063224} expresses that the flux \(\vec J \cdot \vec n\) must be \emph{anti-periodic} at the boundary of \(\cell\): \(\vec x^-\) and \(\vec x^+\) being two homologous points on opposite faces of \(\partial \cell\), \(\vec n^\pm\) denotes the outer unit normal to \(\cell\) at point \(\vec x^\pm\).
\end{problem}

\begin{remark}
The solution $\Phi^\per$ to Problem~\ref{pb:20201012143332} is naturally searched in $H^1_{\rm per}(\cell)$. We observe that $\vec J \in H(\div,\cell)$, and its normal trace $\vec J \cdot \vec n$ (which appears in~\eqref{eq:20200514063224}) is thus well defined (see Appendix~\ref{app:H_div}).
\end{remark}

Owing to its linearity, the solution to Problem~\ref{pb:20201012143332} depends linearly on the sole parameter, namely the constant macroscopic electric field \(\overline{\vec E} \in \reals^d\). In particular, the apparent conductivity \(\tens{\sigma}^\app(\omega,L)\) is defined as the linear mapping between the macroscopic electric field \(\overline{\vec E}\) and the volume average of the electric current \(\langle\vec J\rangle\):
\begin{equation} \label{eq:20200514063314}
  \langle \vec J \rangle = \tens{\sigma}^\app(\omega,L) \ \overline{\vec E},
\end{equation}
where angle brackets refer to volume averages over \(\cell\). The above relation also writes
$$
\frac{1}{|\cell|} \int_\cell \tens{\sigma} \left( \overline{\vec E} - \grad\Phi^\per \right) = \tens{\sigma}^\app(\omega,L) \ \overline{\vec E}.
$$
It should be observed that~\eqref{eq:20200616062027} and the periodicity of \(\Phi^\per\) lead to \(\langle\vec E\rangle = \overline{\vec E}\) and the equality~\eqref{eq:20200514063314} therefore also reads
$$
\langle \vec J \rangle = \tens{\sigma}^\app(\omega,L) \ \langle \vec E \rangle.
$$
We emphasize that the apparent conductivity depends on the realization \(\omega\) (which introduces a random error with respect to $\tens{\sigma}^\eff$) and on the size \(L\) of the SVE (which introduces a bias since, in general, $\ensavg[\tens{\sigma}^\app(\omega,L)] \neq \tens{\sigma}^\eff$; see e.g.~\citet{kani2003}).

For statistically homogeneous, ergodic systems, the \emph{effective} conductivity introduced in the homogenized problem~\eqref{eq:pb_homog} (which, we recall, is defined in terms of the correctors $\Phi$ solution to~\eqref{eq:pb_corr}) is the limit of the above \emph{apparent} conductivity when \(L\) grows to infinity (see e.g.~\citet{sab1992} and~\citet{bourgeat04}):
$$
\tens{\sigma}^\eff = \lim_{L\to+\infty} \tens{\sigma}^\app(\omega,L) \quad \text{almost surely in $\omega$}.
$$
We recall that the realization \(\tens{\sigma}(\cdot,\omega)\) fills the whole space: the apparent conductivity is therefore defined for all \(L\), and the above limit is meaningful. It turns out (this is the manifestation of the law of large numbers) that this limit does not depend on the realization \(\omega\) under consideration. However, in order to obtain a better approximation of $\tens{\sigma}^\eff$, the above formula is often applied as
$$
\tens{\sigma}^\eff = \lim_{L\to+\infty} \ensavg[\tens{\sigma}^\app(\omega,L)].
$$
The above expectation is evaluated as an empirical average. In practice, the typical homogenization workflow is as follows:
\begin{enumerate}
\item generate \(M\) realizations of the microstructure $\tens{\sigma}_1(\cdot,\omega)$, \ldots, $\tens{\sigma}_M(\cdot,\omega)$;
\item \label{item:20200514065157} for a fixed value of \(L\), solve Problem~\ref{pb:20201012143332} for the $M$ realizations $\tens{\sigma}_m(\cdot,\omega)$;
\item compute the apparent conductivities $\tens{\sigma}^\app_m(\omega,L)$ for all $1 \leq m \leq M$;
\item \label{item:20200514065225} estimate the average apparent conductivity
  $$
  \ensavg[\tens{\sigma}^\app(\omega,L)] \simeq \frac{1}{M} \sum_{m=1}^M \tens{\sigma}^\app_m(\omega,L);
  $$
\item possibly repeat for larger values of \(L\).
\end{enumerate}
The above workflow entails three types of numerical errors. In Step~\ref{item:20200514065157}, discretization errors can typically be controlled by the coarseness of the mesh. In Step~\ref{item:20200514065225}, statistical errors are controlled through the number \(M\) of realizations (see below), while finite-size errors can be reduced by increasing the size \(L\) of the SVE. The present article is devoted to statistical errors: it will therefore be assumed that the discretization error is negligible and that the size \(L\) of the SVE is \emph{fixed}. The fixed domain \(\cell\) occupied by the SVE will be referred to as the \emph{unit-cell}, and we will drop any references to its size \(L\): we write \(\tens{\sigma}^\app(\omega)\) rather than \(\tens{\sigma}^\app(\omega,L)\) in the remainder of this article.

\subsection{Overview of the equivalent inclusion method} \label{sec:20201006103812}

The equivalent inclusion method (EIM) was first introduced by~\citet{mosc1975} to compute approximations of the effective response of an assembly of inhomogeneities embedded in a homogeneous matrix. In the present section, we give a brief overview of the variational form of this method, proposed by~\citet{bris2014}.

Note that, following~\citet{eshe1957} and~\citet{mosc1975}, we define an \emph{inclusion} as a bounded domain with same conductivity as the surrounding matrix, subjected to a non-zero electric polarization (to be defined below) and an \emph{inhomogeneity} as a bounded domain with homogeneous conductivity that does not match that of the surrounding matrix.

\subsubsection{The Lippmann--Schwinger equation}

The starting point of the EIM is the Lippmann--Schwinger equation, which is a reformulation of the corrector problem~\eqref{eq:20200616062014}--\eqref{eq:20200514063224} (see~\citet{korr1973,zell1973,kron1974}). This equation requires a so-called \emph{reference material} with homogeneous conductivity \(\tens{\sigma}_0\) (symmetric, positive definite second-order tensor). The reference material being selected, we define the following auxiliary problem:

\begin{problem} \label{pb:20201012145442}
  The \emph{electric polarization} \(\vec P\) being prescribed (as a \(\cell\)-periodic vector field with square integrable components), find the \(\cell\)-periodic electric potential \(\Phi^\per\) such that
  \begin{gather*}
    \div \vec J = 0 \quad \text{in $\cell$}, \\
    \vec J = \tens{\sigma}_0 \, \vec E + \vec P \quad \text{in $\cell$}, \\
    \vec E = -\grad\Phi^\per \quad \text{in $\cell$}, \\
    \vec J^+ \cdot \vec n^+ = - \vec J^- \cdot \vec n^- \quad \text{on $\partial \cell$}.
  \end{gather*}
\end{problem}
For any $\vec P \in (L^2(\cell))^d$, we observe that the solution $\vec J$ to Problem~\ref{pb:20201012145442} belongs to $H(\div,\cell)$, and its normal trace (which appears in the last equation above) is thus well defined (see Appendix~\ref{app:H_div}). Note that we do not request $\vec P$ to have a well-defined normal trace.

Problem~\ref{pb:20201012145442} is linear with respect to the parameter \(\vec P\). The Green operator \(\tens{\Gamma}_0\) is defined as the linear operator that maps the vector field \(\vec P\) onto \(-\vec E\), where \(\vec E\) denotes the unique solution to Problem~\ref{pb:20201012145442}:
\begin{equation} \label{eq:def_Gamma0}
\tens{\Gamma}_0(\vec P) = - \vec E = \grad\Phi^\per.
\end{equation}
In the particular case of isotropic homogeneous reference materials, the conductivity tensor \(\tens{\sigma}_0\) is proportional to the second-order identity tensor \(\tens{I}\), that is \(\tens{\sigma}_0 = \sigma_0 \, \tens{I}\), and the Green operator \(\tens{\Gamma}_0\) has a closed form expression in Fourier space (see~\citet{milt2002}):
\begin{equation} \label{eq:20201006145808}
  \tens{\Gamma}_0(\vec P)(\vec x) = \sum_{\vec n \in \integers^d} \widehat{\tens{\Gamma}}_0 (\vec k_{\vec n}) \ \widetilde{\vec P}_{\vec n} \, \exp\bigl(\I \, \vec k_{\vec n} \cdot \vec x\bigr),
\end{equation}
where $\vec k_{\vec n} = 2\pi\,\vec n / L$, \(\widetilde{\vec P}_{\vec n}\) is the \(\vec n\)-th Fourier coefficient of \(\vec P\) (our convention on Fourier transforms are recalled in Appendix~\ref{sec:20201006143813}), and where the second-order tensor $\widehat{\tens{\Gamma}}_0(\vec k)$ is given, for any \(\vec k \in \reals^d\), by
\begin{equation} \label{eq:20200722092039}
  \widehat{\tens{\Gamma}}_0(\vec k)=
  \begin{cases}
    \displaystyle \frac{1}{\sigma_0} \, \frac{\vec k\otimes\vec k}{| \vec k |^2} & \text{for $\vec k \neq \vec 0$},\\
    \tens{0} & \text{for $\vec k = \vec 0$}.
  \end{cases}
\end{equation}
The Green operator \(\tens{\Gamma}_0\) being thus defined by~\eqref{eq:def_Gamma0}, it can be shown that Problem~\ref{pb:20201012143332} is equivalent to the Lippmann--Sch\-winger equation
\begin{equation} \label{eq:20200617061810}
  \vec E + \tens{\Gamma}_0\bigl[ \bigl(\tens{\sigma}-\tens{\sigma}_0\bigr) \, \vec E \bigr] = \overline{\vec E},
\end{equation}
where the unknown is the electric field \(\vec E \in (L^2(\cell))^d \). In turn, introducing the polarization \(\vec P = \bigl(\tens{\sigma}-\tens{\sigma}_0\bigr) \, \vec E\), the above equation can be transformed as
\begin{equation} \label{eq:20200616063758}
  \bigl(\tens{\sigma}-\tens{\sigma}_0\bigr)^{-1} \, \vec P + \tens{\Gamma}_0(\vec P) = \overline{\vec E},
\end{equation}
and the apparent conductivity defined by~\eqref{eq:20200514063314} is obtained from the average polarization by
\begin{equation} \label{eq:20200617083659}
  \begin{aligned}[b]
    \tens{\sigma}^\app \, \overline{\vec E}
    &= \langle \vec J \rangle
    = \langle \tens{\sigma} \, \vec E \rangle
    = \langle \tens{\sigma}_0 \, \vec E + \vec P \rangle \\
    &= \tens{\sigma}_0 \, \langle \vec E \rangle + \langle \vec P \rangle
    = \tens{\sigma}_0 \, \overline{\vec E} + \langle \vec P \rangle.
  \end{aligned}
\end{equation}
Solving~\eqref{eq:20200616063758} is equivalent to solving Problem~\ref{pb:20201012143332}. In the following, we turn to discretization approaches for~\eqref{eq:20200616063758}.

\subsubsection{The equivalent inclusion method} \label{sec:eim_method}

The equivalent inclusion method of~\citet{mosc1975} delivers an approximate solution to the Lippmann--Sch\-winger equation~\eqref{eq:20200616063758} for materials made of well-defined individual phases.

We consider that the heterogeneous material under investigation is made of \(N+1\) phases \(\Omega_1, \ldots, \Omega_{N+1} \subset \reals^d\) that form a partition of \(\reals^d\). ``Phase'' should be understood here in a loose sense: each \(\Omega_\alpha\) has uniform conductivity \(\tens{\sigma}_\alpha\). As an example, we refer to the microstructure shown on Fig.~\ref{fig:20220727151638} below, which contains $N$ inhomogeneities in $\cell$. Each inhomogeneity (along with its periodic images in $\reals^d$) is considered to be a phase. Phase \(N+1\) will be particularized, and referred to as the ``matrix''.

We introduce the indicator function \(\chi_\alpha \colon \reals^d \longrightarrow \{0,1\}\) of \(\Omega_\alpha\) and the volume fraction of phase \(\alpha\) in the unit-cell \(\cell\), \(f_\alpha = \langle \chi_\alpha \rangle= | \Omega_\alpha \cap \cell | / | \cell |\).

In the most simple form of the EIM, a piecewise constant approximate solution of~\eqref{eq:20200616063758} is constructed and \emph{the reference material coincides with the matrix}: \(\tens{\sigma}_0 = \tens{\sigma}_{N+1}\). The polarization \(\vec P\) therefore vanishes over \(\Omega_{N+1}\) (since \(\vec P_{N+1} = (\tens{\sigma}_{N+1} - \tens{\sigma}_0) \, \vec E\)). Based on this approximation space, discretization of the Lippmann--Schwinger equation initially relied on collocation~\citep{mosc1975}. More recently, \citet{bris2014} proposed a Galerkin discretization of the variational form of this equation. As briefly outlined in Appendix~\ref{sec:20201013074424}, Equation~\eqref{eq:20200616063758} reduces after discretization to the following linear system with unknowns \(\vec P_1, \ldots,\vec P_N \in \reals^d\) (the constant polarizations in each phase; recall that \(\vec P_{N+1} = \vec 0\) in the matrix phase): for any $1 \leq \alpha \leq N$,
\begin{equation} \label{eq:20200629090809}
  \bigl(\tens{\sigma}_\alpha-\tens{\sigma}_0\bigr)^{-1} \, \vec P_\alpha + \sum_{\beta=1}^N \tens{\Gamma}_{\alpha\beta} \ \vec P_\beta = \overline{\vec E}.
\end{equation}
We are thus left with solving a linear system with \(Nd\) scalar unknowns. In~\eqref{eq:20200629090809}, the so-called second-order \emph{influence tensors} \(\tens{\Gamma}_{\alpha\beta}\) are defined by how they operate on (constant) vectors: for any \(\vec P\in\reals^d\),
\begin{equation} \label{eq:20200721095808}
\tens{\Gamma}_{\alpha\beta} \, \vec P = | \Omega_\alpha \cap \cell |^{-1} \int_{\Omega_\alpha \cap \cell} \tens{\Gamma}_0(\chi_\beta \, \vec P)(\vec x) \, \D \vec x.
\end{equation}
The quantity \(-\tens{\Gamma}_{\alpha\beta} \vec P\) is the average electric field (in a periodic setting) induced in phase \(\alpha\) by a polarization which is constant and equal to \(\vec P\) in the phase \(\beta\) and vanishes everywhere else. Note that, from the point of view of~\eqref{eq:20200721095808}, \(\Omega_\alpha\) and \(\Omega_\beta\) should be seen as \emph{inclusions} embedded in a homogeneous matrix with conductivity \(\tens{\sigma}_0\) (as discussed at the beginning of Sec.~\ref{sec:20201006103812}).

The EIM approximation of the apparent conductivity is deduced from the solution to~\eqref{eq:20200629090809} by the following identity, which directly follows from~\eqref{eq:20200617083659}:
\begin{equation} \label{eq:20201013075616}
  \tens{\sigma}^\EIM \, \overline{\vec E} = \tens{\sigma}_0 \, \overline{\vec E} + \sum_{\alpha=1}^N f_\alpha \, \vec P_\alpha.
\end{equation}
As usual in computational homogenization, the full determination of the conductivity tensor \(\tens{\sigma}^\EIM\) requires~\eqref{eq:20200629090809} to be solved and~\eqref{eq:20201013075616} to be evaluated for \(d\) linearly independent values of the constant macroscopic electric field \(\overline{\vec E} \in \reals^d\). Since we consider a random setting and a bounded SVE, the EIM approximation $\tens{\sigma}^\EIM$ of course depends on \(\omega\).

\begin{remark}[Assembly of the linear system]
  The linear system \(Ax=b\) corresponding to~\eqref{eq:20200629090809} is assembled by blocks:
  \begin{gather*}
    A_{(\alpha-1)d+i, (\beta-1)d+j} = \vec e_i \cdot \bigl[\delta_{\alpha\beta} \, \bigl(\tens{\sigma}_\alpha-\tens{\sigma}_0\bigr)^{-1} + \tens{\Gamma}_{\alpha\beta} \bigr] \, \vec e_j, \\
    b_{(\alpha-1)d+i} = \overline{\vec E} \cdot \vec e_i,
  \end{gather*}
  where \(1 \leq \alpha, \beta \leq N\) and \(1 \leq i, j \leq d\). The solution \(x\) to the linear system \(Ax=b\) delivers the polarizations in each inclusion: $\vec P_\alpha \cdot \vec e_i = x_{(\alpha-1)d+i}$.
\end{remark}

As pointed out in the introduction, there is a strong link between the EIM method and the Hashin-Shtrikman variational principle (as introduced in~\citet{hash1962}) and the Hashin-Shtrikman bounds (proposed in~\citet{hash1962a}). The reformulation of the corrector problem~\ref{pb:20201012143332} as the Lippmann--Sch\-winger equation~\eqref{eq:20200617061810}, which is next recast as the equation~\eqref{eq:20200616063758} on the polarization $\vec P$, is at the basis of the Hashin-Shtrikman variational principle. The Hashin-Shtrikman bounds are obtained by taking $\tens{\sigma}_0 = \tens{\sigma}_{N+1}$ (the reference material coincides with the matrix) and searching an approximate solution to~\eqref{eq:20200616063758} in the form of a polarization $\vec P$ which is constant in each inclusion, the constant being {\em the same} for all the inclusions and all the realizations. Here, we also take $\tens{\sigma}_0 = \tens{\sigma}_{N+1}$ but we search for an approximate solution to~\eqref{eq:20200616063758} in the form of a polarization $\vec P$ which is constant in each inclusion, the constant being a priori {\em different} for each inclusion and for each realization.

\subsubsection{Fourier expansion of the influence tensors}

To compute the above approximation $\tens{\sigma}^\EIM$, we need to compute beforehand the influence tensors $\tens{\Gamma}_{\alpha\beta}$. We underline in this section that this computation is actually challenging. However, in the present context where we use the EIM method only as a surrogate model (and not as an approximation of the apparent conductivity $\tens{\sigma}^\app$), it turns out, as explained in Sec.~\ref{sec:20220729143106}, that we can circumvent this difficult and adopt an alternative strategy.

The standard approach to compute $\tens{\Gamma}_{\alpha\beta}$ is as follows. The following expression of the influence tensors is derived in Appendix~\ref{sec:20201006153018}:
\begin{equation} \label{eq:20220705185558}
  f_\alpha \, \tens{\Gamma}_{\alpha\beta} = \sum_{\vec n \in \integers^d} \widetilde{\chi}_{\alpha, \vec n}^\ast \ \widetilde{\chi}_{\beta, \vec n} \ \widehat{\tens{\Gamma}}_0(\vec k_{\vec n}),
\end{equation}
where $\widetilde{\chi}_{\alpha, \vec n}^\ast$ is the complex conjugate of the Fourier coefficient $\widetilde{\chi}_{\alpha, \vec n}$ of the indicator function $\chi_\alpha$ of the phase $\alpha$,
\begin{equation} \label{eq:20220705190239}
  \widetilde{\chi}_{\alpha, \vec n} = \frac{1}{| \cell |} \int_\cell \chi_\alpha(\vec x) \, \exp\bigl(-\I \, \vec k_{\vec n} \cdot \vec x \bigr) \, \D \vec x.
\end{equation}

In the particular case when $d=2$ and when the phase $\alpha$ occupies a disk \(\Omega_\alpha\) of center \(\vec x_\alpha\) and of radius \(a_\alpha\), we have, as shown in Appendix~\ref{sec:20201006153018}, that
\begin{equation} \label{eq:20220706162136}
  \widetilde{\chi}_{\alpha, \vec n} = f_\alpha \, F(k_{\vec n} \, a_\alpha) \, \E^{-\I \, \vec k_{\vec n} \cdot \vec x_\alpha} \quad \text{with} \quad F(\xi) = 2 \besselj_1(\xi) / \xi,
\end{equation}
where \(\besselj_1\) denotes the Bessel function of the first kind and \(k_{\vec n} = | \vec k_{\vec n} |\). Inserting~\eqref{eq:20220706162136} into~\eqref{eq:20220705185558}, we find, with \(\vec r_{\alpha \beta} = \vec x_\beta - \vec x_\alpha\), that
$$
\tens{\Gamma}_{\alpha\beta} = f_\beta \sum_{\vec n \in \integers^2} F(k_{\vec n} \, a_\alpha) \, F(k_{\vec n} \, a_\beta) \, \E^{-\I \, \vec k_{\vec n} \cdot \vec r_{\alpha \beta}} \ \widehat{\tens{\Gamma}}_0(\vec k_{\vec n}).
$$

In dimension $d=2$, and when the phases occupy a regular shape, the general term of the series~\eqref{eq:20220705185558} behaves as \(n^{-2}\)~\citep{deby1957}. In other words, convergence is very slow, which makes the numerical evaluation of the influence tensors \(\tens{\Gamma}_{\alpha\beta}\) difficult. Several acceleration techniques have been proposed for an accurate evaluation of these coefficients~\citep{to2016}. We do not detail them here since, as pointed out above, in our particular context, we can adopt an alternative strategy, as described in Sec.\ref{sec:20220729143106}.

\section{EIM-based variance reduction} \label{sec:20220801101220}

We have shown in Sec.~\ref{sec:20220801153432} how the EIM can deliver an approximation of the apparent conductivity. This estimate, given by~\eqref{eq:20201013075616}, is generally much cheaper to compute than a full-field estimate, since it involves \(Nd\) scalar unkowns only (where $N$ is the number of phases when ignoring the matrix phase). Hoping for a good correlation between the true apparent conductivity $\tens{\sigma}^\app(\omega)$ and its EIM estimate $\tens{\sigma}^\EIM(\omega)$, we now discuss how to use the EIM estimate as a surrogate model for variance reduction.

\subsection{Outline of the strategy} \label{sec:20210915121856}

Given a multi-phase material, our goal is to estimate the ensemble average \(\ensavg[\tens{\sigma}^\app(\omega)]\) (which is an average over all realizations of the microstructure). We recall that we are not concerned with finite-size effects in this article (the size \(L\) of the SVE is therefore fixed).

As discussed in Sec.~\ref{sec:20200514070529}, the ensemble average \(\ensavg[\tens{\sigma}^\app(\omega)]\) is in practice approximated by an empirical average over \(M\) realizations of the apparent conductivities \(\tens{\sigma}^\app_m(\omega)\), $1 \leq m \leq M$, themselves evaluated by means of a full-field simulation (typically, using the finite element method~--~FEM):
$$
\ensavg(\tens{\sigma}^\app) \simeq \frac{1}{M} \sum_{m=1}^M \tens{\sigma}^\FEM_m(\omega),
$$
where \(\tens{\sigma}^\FEM_m(\omega)\) denotes the finite element numerical approximation of the apparent conductivity of the $m$-th realization.

In order to increase the reliability of this estimate, we apply the variance reduction strategy introduced in Sec.~\ref{sec:20201013092415} with the EIM estimate of the apparent conductivity as surrogate model. More precisely, in order to reduce the variance on the full-field estimate of \emph{one component} of the apparent conductivity, we use the EIM estimate of the \emph{same component} as a surrogate model.

\begin{remark}[Ensemble average of the surrogate model] \label{rem:expect_unknown2}
  Note that, for any component $1 \leq i,j \leq d$ of the conductivity tensor, the ensemble average $\ensavg(\sigma_{ij}^\EIM)$ is not known. In the vein of Remark~\ref{rem:expect_unknown}, a Monte--Carlo estimate is therefore used here, with a number \({\cal M}\) of realizations that is significantly larger than the number $M$ of full-field simulations. Owing to the comparatively low cost of the equivalent inclusion method, these extra numerical computations do not jeopardize the overall efficiency of the proposed strategy.
\end{remark}

\begin{remark}
  It is in theory possible to use a tensor version of the variance reduction method described in Sec.~\ref{sec:20201013092415}, where \emph{any component} of \(\tens{\sigma}^\FEM\) is cross-correlated with \emph{all components} of \(\tens{\sigma}^\EIM\). We have implemented this approach and observed that it led to negligible improvements. The present article is therefore devoted to the case where only the component \(\sigma_{ij}^\EIM\) is used to improve the computation on $\ensavg(\sigma_{ij}^\FEM)$.
\end{remark}

To summarize the above discussion, the proposed workflow is outlined as follows (as recalled above, the size \(L\) of the SVE is \emph{fixed}):
\begin{enumerate}
\item Generate \({\cal M} \gg M\) realizations of the microstructure.
\item \label{item:20220707093953} Perform FEM simulations on the first \(M\) microstructures. This step delivers the apparent conductivities \(\tens{\sigma}^\FEM_1(\omega)\), \dots, \(\tens{\sigma}^\FEM_M(\omega)\).
\item \label{item:20220707093754} Perform EIM simulations on all ${\cal M}$ microstructures. For each realization, this step requires to assemble and solve the linear system~\eqref{eq:20200629090809}; the EIM approximations \(\tens{\sigma}^\EIM_1(\omega)\), \dots, \(\tens{\sigma}^\EIM_{\cal M}(\omega)\) are obtained by~\eqref{eq:20201013075616}.
\item \label{item:20220707094018} Use all the results of Step~\ref{item:20220707093754} to compute an accurate estimate of \(\ensavg(\tens{\sigma}^\EIM)\), in the vein of Remarks~\ref{rem:expect_unknown} and~\ref{rem:expect_unknown2}.
\item For any \(1 \leq i, j \leq d\), use the results from Steps~\ref{item:20220707093953}, \ref{item:20220707093754} and~\ref{item:20220707094018} to apply the variance reduction technique described in Sec.~\ref{sec:20201013092415} to \(X = \sigma_{ij}^\FEM\) and \(Y = \sigma_{ij}^\EIM\).
\end{enumerate}

Of course, for this strategy to be viable, it is essential that Step~\ref{item:20220707093754} be performed in negligible time. The linear system to be solved is small. However, its assembly is complex, because of the slow convergence of the Fourier series~\eqref{eq:20220705185558} that defines the influence tensors. This problem is addressed in the next section.

\subsection{Efficient evaluation of the influence tensors} \label{sec:20220729143106}

The crucial observation is that we do not need an exact evaluation of the influence tensors $\tens{\Gamma}_{\alpha\beta}$. Indeed, as a surrogate model, \(\tens{\sigma}^\EIM\) is not required to be an accurate estimate of the true conductivity \(\tens{\sigma}^\app\). The only requirement is that both random variables \(\tens{\sigma}^\FEM\) and \(\tens{\sigma}^\EIM\) be \emph{correlated}. In other words, alterations of the EIM linear system are possible, as long as such alterations do not pollute the correlations between FEM and EIM simulations.

A natural idea is to replace the infinite series~\eqref{eq:20220705185558} with the truncated sum
$$
f_\alpha \, \tens{\Gamma}_{\alpha\beta} \simeq \sum_{-P \leq n_1, \ldots, n_d \leq P} \widetilde{\chi}_{\alpha, \vec n}^\ast \ \widetilde{\chi}_{\beta, \vec n} \ \widehat{\tens{\Gamma}}_0(\vec k_{\vec n}),
$$
where \(P\) is sufficiently small to keep the cost of this \(d\)-dimen\-sional sum to a minimum (typically, \(P\leq 3\)). Our numerical experiments (not reproduced here) show that such an approach does not lead to satisfactory variance reduction, which probably means that the neglected high-frequency terms have an important correlation content.

A different approach was suggested by the recent work of~\citet{zece2021}. Within the framework of ``FFT-based homogenization methods''~\citep{moul1994, moul1998}, they considered series similar to~\eqref{eq:20220705185558}, \(\chi_\alpha\) and \(\chi_\beta\) being the indicator functions of two ``voxels''. Using the Poisson formula, they first transformed the series in Fourier space into a series in the real-space, that they then truncate. We propose the same transformation based on the Poisson formula. However, unlike~\citet{zece2021}, we will keep a minimum number of terms in the truncated sum. Our goal in this section is to provide an insight into the physical meaning of the transformation of~\citet{zece2021} and also propose a correction to this formula.

The microstructure being periodic, it is interesting to define, for any phase \(1 \leq \alpha \leq N\), the \emph{reference} domain \(\Omega_\alpha^0\), such that the whole phase \(\Omega_\alpha\) is covered by periodic copies of \(\vec x_\alpha + \Omega_\alpha^0\), \(\vec x_\alpha\) being a fixed position in the unit-cell \(\cell\). For example, in the case of circular inclusions, \(\Omega_\alpha^0\) is the disk centered at the origin (with radius \(a_\alpha\)) and \(\vec x_\alpha\) is the center of the sole periodic image the center of which lies in the unit-cell \(\cell\) (see Fig.~\ref{fig:20220729144521}). From the above definitions, we have (`+' referring to translation)
$$
\Omega_\alpha = \bigcup_{\vec n \in \integers^d} \bigl[ \bigl( \vec x_\alpha + \vec n \, L \bigr) + \Omega_\alpha^0 \bigr].
$$

\setlength{\sbL}{2.4cm}

\newlength{\sbxA}
\setlength{\sbxA}{-0.42\sbL}
\newlength{\sbyA}
\setlength{\sbyA}{-0.44\sbL}
\newlength{\sbrA}
\setlength{\sbrA}{4mm}

\begin{figure}
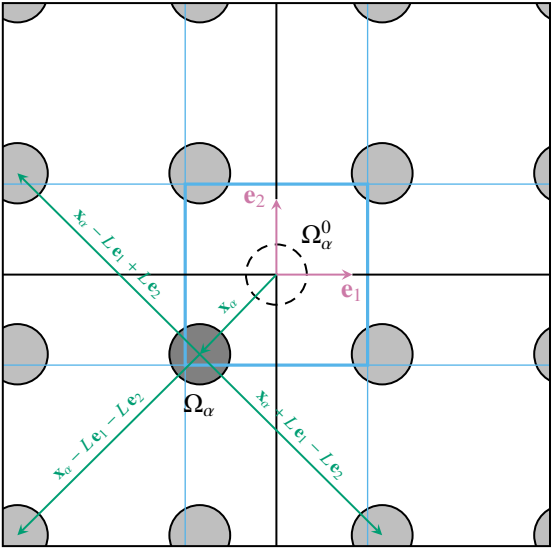

  \centering
  \begin{pgfpicture}
    \pgfslopedattimetrue
    \pgfsetlinewidth{\pgfcadnormal}

    \begin{pgfscope}
      \pgfpathrectangle{\pgfpoint{-1.5\sbL}{-1.5\sbL}}{\pgfpoint{3\sbL}{3\sbL}}
      \pgfusepath{clip}

      \pgfsetdash{{1.5mm}{1mm}}{0pt}
      \pgfpathcircle{\pgfpointorigin}{\sbrA}
      \pgfusepath{stroke}
      \pgfsetdash{}{0pt}

      \pgftransformshift{\pgfpoint{\sbxA}{\sbyA}}

      \pgfsetfillcolor{gray}
      \pgfpathcircle{\pgfpointorigin}{\sbrA}
      \pgfusepath{stroke, fill}

      \pgfsetfillcolor{lightgray}
      \pgfpathcircle{\pgfpoint{-\sbL}{-\sbL}}{\sbrA}
      \pgfpathcircle{\pgfpoint{     0}{-\sbL}}{\sbrA}
      \pgfpathcircle{\pgfpoint{ \sbL}{-\sbL}}{\sbrA}
      \pgfpathcircle{\pgfpoint{2\sbL}{-\sbL}}{\sbrA}
      \pgfpathcircle{\pgfpoint{-\sbL}{    0}}{\sbrA}
      \pgfpathcircle{\pgfpoint{ \sbL}{    0}}{\sbrA}
      \pgfpathcircle{\pgfpoint{2\sbL}{    0}}{\sbrA}
      \pgfpathcircle{\pgfpoint{-\sbL}{ \sbL}}{\sbrA}
      \pgfpathcircle{\pgfpoint{    0}{ \sbL}}{\sbrA}
      \pgfpathcircle{\pgfpoint{ \sbL}{ \sbL}}{\sbrA}
      \pgfpathcircle{\pgfpoint{2\sbL}{ \sbL}}{\sbrA}
      \pgfpathcircle{\pgfpoint{-\sbL}{2\sbL}}{\sbrA}
      \pgfpathcircle{\pgfpoint{    0}{2\sbL}}{\sbrA}
      \pgfpathcircle{\pgfpoint{ \sbL}{2\sbL}}{\sbrA}
      \pgfpathcircle{\pgfpoint{2\sbL}{2\sbL}}{\sbrA}

      \pgfusepath{stroke, fill}
    \end{pgfscope}

    \begin{pgfscope}
      \pgfsetstrokecolor{sblightblue}
      \pgfsetlinewidth{\pgfcadthin}
      \pgfpathmoveto{\pgfpoint{-0.5\sbL}{-1.5\sbL}}
      \pgfpathlineto{\pgfpoint{-0.5\sbL}{1.5\sbL}}
      \pgfpathmoveto{\pgfpoint{0.5\sbL}{-1.5\sbL}}
      \pgfpathlineto{\pgfpoint{0.5\sbL}{1.5\sbL}}
      \pgfpathmoveto{\pgfpoint{-1.5\sbL}{-0.5\sbL}}
      \pgfpathlineto{\pgfpoint{1.5\sbL}{-0.5\sbL}}
      \pgfpathmoveto{\pgfpoint{-1.5\sbL}{0.5\sbL}}
      \pgfpathlineto{\pgfpoint{1.5\sbL}{0.5\sbL}}
      \pgfusepath{stroke}

      \pgfsetlinewidth{\pgfcadthick}
      \pgfpathrectangle{\pgfpoint{-0.5\sbL}{-0.5\sbL}}{\pgfpoint{\sbL}{\sbL}}
      \pgfusepath{stroke}
    \end{pgfscope}

    \pgfpathmoveto{\pgfpoint{-1.5\sbL}{0}}
    \pgfpathlineto{\pgfpointorigin}
    \pgfpathlineto{\pgfpoint{0}{-1.5\sbL}}
    \pgfpathmoveto{\pgfpoint{\sbunitvectorlength}{0}}
    \pgfpathlineto{\pgfpoint{1.5\sbL}{0}}
    \pgfpathmoveto{\pgfpoint{0}{\sbunitvectorlength}}
    \pgfpathlineto{\pgfpoint{0}{ 1.5\sbL}}
    \pgfusepath{stroke}

    \begin{pgfscope}
      \color{sbpurple}
      \pgfsetstrokecolor{sbpurple}
      \pgfsetfillcolor{sbpurple}
      \pgfsetarrows{stealth-stealth}
      \pgfpathmoveto{\pgfpoint{0}{\sbunitvectorlength}}
      \pgfpathlineto{\pgfpointorigin}
      \pgfpathlineto{\pgfpoint{\sbunitvectorlength}{0}}
      \pgfusepath{stroke}
      \begin{pgfscope}
        \pgftransformshift{\pgfpoint{\sbunitvectorlength}{0}}
        \pgfnode{rectangle}{north}{\(\vec e_1\)}{}{\pgfusepath{}}
      \end{pgfscope}
      \begin{pgfscope}
        \pgftransformshift{\pgfpoint{0}{\sbunitvectorlength}}
        \pgfnode{rectangle}{east}{\(\vec e_2\)}{}{\pgfusepath{}}
      \end{pgfscope}
    \end{pgfscope}

    \begin{pgfscope}
      \scriptsize
      \pgfsetarrowsend{stealth}
      \pgfsetstrokecolor{sbgreen}

      \pgfcaddecoratedline{0.5}{\pgfpointorigin}{\pgfpoint{\sbxA}{\sbyA}}{south}{\(\color{sbgreen}\vec x_\alpha\)}

      \pgftransformshift{\pgfpoint{\sbxA}{\sbyA}}

      \begin{pgfscope}
        \normalsize
        \pgftransformshift{\pgfpoint{0}{-\sbrA}}
        \pgfnode{rectangle}{north}{\(\Omega_\alpha\)}{}{\pgfusepath{}}
      \end{pgfscope}

      \pgfcaddecoratedline{0.5}{\pgfpointorigin}{\pgfpoint{-\sbL}{-\sbL}}{south}{\(\color{sbgreen}\vec x_\alpha-L\,\vec e_1-L\,\vec e_2\)}
      \pgfcaddecoratedline{0.5}{\pgfpointorigin}{\pgfpoint{-\sbL}{\sbL}}{south}{\(\color{sbgreen}\vec x_\alpha-L\,\vec e_1+L\,\vec e_2\)}
      \pgfcaddecoratedline{0.5}{\pgfpointorigin}{\pgfpoint{\sbL}{-\sbL}}{south}{\(\color{sbgreen}\vec x_\alpha+L\,\vec e_1-L\,\vec e_2\)}
    \end{pgfscope}

    \begin{pgfscope}
      \pgftransformshift{\pgfpoint{.5\sbrA}{.5\sbrA}}
      \pgfnode{rectangle}{south west}{\(\Omega_\alpha^0\)}{}{\pgfusepath{}}
    \end{pgfscope}

    \begin{pgfscope}
      \pgfpathrectangle{\pgfpoint{-1.5\sbL}{-1.5\sbL}}{\pgfpoint{3\sbL}{3\sbL}}
      \pgfusepath{stroke}
    \end{pgfscope}
  \end{pgfpicture}
  \caption{Unit-cell $\cell$ (solid blue line) and reference domain \(\Omega_\alpha^0\) (black dashed line). The domain \(\Omega_\alpha\) occupied by the phase $\alpha$ is the union of the principal image (centered at \(\vec x_\alpha\), dark grey; by definition, \(\vec x_\alpha \in \cell\); note that, in the case represented on the figure, the principal image $\vec x_\alpha + \Omega_\alpha^0$ is not a subset of $\cell$) and of its periodic images (light gray). \label{fig:20220729144521}}
\end{figure}

It is recalled (see~\eqref{eq:20200721095808}) that, up to a sign change, \(\tens{\Gamma}_{\alpha\beta}\) measures the average electric field induced in phase \(\Omega_\alpha\) by a constant polarization of phase \(\beta\). Owing to the periodicity of the microstructure, averaging over phase \(\Omega_\alpha\) is equivalent to averaging over the principal image \(\vec x_\alpha + \Omega_\alpha^0\). Besides, \(\Omega_\beta\) is made of the \emph{principal image} \(\vec x_\beta + \Omega_\beta^0\) and all its periodic images. The tensor \(\tens{\Gamma}_{\alpha\beta}\) therefore measures the average electric field induced in the fundamental image \(\vec x_\alpha + \Omega_\alpha^0\) by a constant polarization of all periodic images of \(\vec x_\beta + \Omega_\beta^0\). In short, \(\tens{\Gamma}_{\alpha\beta}\) measures the interaction between \(\vec x_\alpha + \Omega_\alpha^0\) and all periodic images of \(\vec x_\beta + \Omega_\beta^0\).

Since the conductivity problem under consideration is linear, these interactions are \emph{additive}, and we can focus on the interaction between \(\vec x_\alpha + \Omega_\alpha^0\) and \emph{one} periodic image \(\Omega_\beta^1 = \vec x_\beta + \vec n_1 \, L + \Omega_\beta^0\) (for a given $\vec n_1 \in \integers^d$) of \(\Omega_\beta^0\). By invariance upon translation, we focus on the interaction between \(\Omega_\alpha^0\) and \(\Omega_\beta^1 = \vec r + \Omega_\beta^0\), where \(\vec r = \vec x_\beta - \vec x_\alpha + \vec n_1 \, L\).

Intuition suggests that we are no longer working with periodic boundary conditions. Rather, \(\Omega_\alpha^0\) and \(\Omega_\beta^1\) are embedded in the infinite space \(\reals^d\), and boundary conditions are applied at infinity (to ensure square integrability of the electric field). We then consider the following problem, for a constant vector $\vec P_\beta \in \reals^d$:
\begin{gather}
  \label{eq:edp_espace_infini1}
  \div \vec J = 0 \quad \text{in $\reals^d$}, \\
  \vec J = \tens{\sigma}_0 \vec E + \vec P_\beta \quad \text{in $\Omega_\beta^1$}, \\
  \vec J = \tens{\sigma}_0 \vec E \quad \text{in $\reals^d \setminus \Omega_\beta^1$}, \\
  \label{eq:edp_espace_infini4}
  \vec E = - \grad\Phi \quad \text{in $\reals^d$},
\end{gather}
where $\Phi \in L^2_{\rm loc}(\reals^d)$ and $\vec E \in (L^2(\reals^d))^d$, and we are interested in the average electric field over \(\Omega_\alpha^0\). In other words, in an infinite medium with conductivity \(\tens{\sigma}_0\), we apply the polarization \(\vec P_\beta\) to the domain \(\Omega_\beta^1 = \vec r + \Omega_\beta^0\), and measure the average induced electric field over \(\Omega_\alpha^0\). By definition, the influence tensor \(\tens{\Gamma}_{\alpha\beta}^\infty(\vec r)\) maps the polarization \(\vec P_\beta\) onto the opposite of the average electric field:
\begin{equation} \label{eq:def_Gamma_infini}
\tens{\Gamma}_{\alpha\beta}^\infty(\vec r) \, \vec P_\beta = -| \Omega_\alpha^0 |^{-1} \int_{\Omega_\alpha^0} \vec E,
\end{equation}
where $\vec E$ is the solution to~\eqref{eq:edp_espace_infini1}--\eqref{eq:edp_espace_infini4}. The influence tensor $\tens{\Gamma}_{\alpha\beta}^\infty(\vec r)$ of course depends on the geometry of the reference domains $\Omega_\alpha^0$ and $\Omega_\beta^0$, and on the vector $\vec r = \vec x_\beta - \vec x_\alpha + \vec n_1 \, L$.

As an example, in Appendix~\ref{sec:20201007151753}, we consider the two-dimen\-sional case, for isotropic homogeneous reference materials ($\tens{\sigma}_0 = \sigma_0 \, \tens{I}$) and when the phases are made of circular inclusions. We denote $a_\beta$ the radius of the inclusions of phase $\beta$. In this case, we show (see~\eqref{eq:utile_pour_trace}) that
\begin{equation} \label{eq:20201007151853}
  \tens{\Gamma}_{\alpha\beta}^\infty(\vec r) = \frac{a_\beta^2}{2 \sigma_0 | \vec r |^2} \, \biggl( \tens{I} - 2 \frac{\vec r \otimes \vec r}{| \vec r |^2} \biggr),
\end{equation}
where $\vec r$ is then the vector going from the center of $\Omega_\alpha^0$ to the center of $\Omega_\beta^1$.

Note that the only difference between \(\tens{\Gamma}_{\alpha\beta}\) and \(\tens{\Gamma}_{\alpha\beta}^\infty\) lies in the boundary conditions (Problem~\ref{pb:20201012145442} is complemented with periodic boundary conditions, while Problem~\eqref{eq:edp_espace_infini1}--\eqref{eq:edp_espace_infini4} is complemented with the ``boundary condition'' $\vec E \in (L^2(\reals^d))^d$).

The above discussion leads to propose the following formula:
\begin{equation} \label{eq:20220707152227}
  \tens{\Gamma}_{\alpha\beta} \stackrel{?}{=} \sum_{\vec n \in \integers^d} \tens{\Gamma}_{\alpha\beta}^\infty(\vec x_\beta - \vec x_\alpha - \vec n \, L),
\end{equation}
which expresses the fact that \(\tens{\Gamma}_{\alpha\beta}\) measures the (additive) interactions between \(\vec x_\alpha + \Omega_\alpha^0\) and all periodic images of \(\vec x_\beta + \Omega_\beta^0\) (as measured by \(\tens{\Gamma}_{\alpha\beta}^\infty\)).

This is, in essence, the formula intially proposed by~\citet{zece2021} (based on the Poisson summation formula). Note that a similar formula was also proposed by~\citet{moli1996}. However, \citet{zece2021} observed that this sum is not absolutely convergent. It is therefore at best conditionally convergent and the summation scheme must be defined precisely. \citet{zece2021} adopted the following scheme:
\begin{equation} \label{eq:20220721131501}
  \tens{\Gamma}_{\alpha\beta} \stackrel{?}{=} \lim_{P \to +\infty} \sum_{-P\leq n_1, \ldots n_d \leq P} \tens{\Gamma}_{\alpha\beta}^\infty(\vec x_\beta - \vec x_\alpha - \vec n \, L).
\end{equation}

While appealing because of its reasonable physical meaning, the above formula cannot be correct. Consider indeed the case of isotropic reference materials, in dimension $d=2$ and for circular inclusions. It is shown in Appendice~\ref{sec:20201006153018} (see~\eqref{eq:20220707143902}) that, for any \(\alpha \neq \beta\),
$$
\tr \tens{\Gamma}_{\alpha\beta} = -\sigma_0^{-1} \, f_\beta,
$$
while we deduce from~\eqref{eq:20201007151853} that, for any \(\alpha \neq \beta\),
$$
\tr \tens{\Gamma}_{\alpha\beta}^\infty = 0.
$$
Equation~\eqref{eq:20220721131501} is inconsistent with the two above relations. Instead, we (heuristically) postulate the following correction (for the two-dimensional case of disks and isotropic reference materials):
\begin{equation} \label{eq:20220719170815}
  \tens{\Gamma}_{\alpha\beta} = \lim_{P \to +\infty} \sum_{-P\leq n_1, \ldots n_d \leq P} \tens{\Gamma}_{\alpha\beta}^\infty(\vec x_\beta - \vec x_\alpha - \vec n \, L) - \frac{f_\beta}{2\sigma_0} \, \tens{I},
\end{equation}
which ensures that the traces of both sides coincide. We have not been able to give a mathematical proof of~\eqref{eq:20220719170815}. However, a numerical validation is proposed in Appendix~\ref{sec:20220719172106}. Besides, it is shown below that this correction leads to the FEM and EIM results being satisfactorily correlated.

In practice, for the applications presented below, we use a truncated form of~\eqref{eq:20220719170815}:
$$
\tens{\Gamma}_{\alpha\beta} \simeq \sum_{-P\leq n_1, \ldots n_d \leq P} \tens{\Gamma}_{\alpha\beta}^\infty(\vec x_\beta - \vec x_\alpha - \vec n \, L) - \frac{f_\beta}{2\sigma_0} \, \tens{I},
$$
with a low value of \(P\) (typically \(P=1, 2, 3\)). Unless otherwise stated, we set \(P=2\) in Sec.~\ref{sec:20220727145908}. Note also that we have considered the approximation
$$\tens{\Gamma}_{\alpha\beta} \simeq \sum_{-P\leq n_1, \ldots n_d \leq P} \tens{\Gamma}_{\alpha\beta}^\infty(\vec x_\beta - \vec x_\alpha - \vec n \, L),$$
which leads to poor results (not reproduced here) in terms of variance reduction efficiency.

\begin{remark}
  The additional term introduced in~\eqref{eq:20220719170815} is certainly \emph{not universal}. Indeed, this term being isotropic, it is questionable that it should apply to situations where inclusions are not circular and/or the unit-cell is not a square. We leave the study of such general settings to future works. 
\end{remark}

\begin{remark}
  \citet{zece2021} showed that~\eqref{eq:20220707152227} delivers good results for the numerical applications they considered. At this stage, the reason for this is unclear to us. 
\end{remark}

The workflow outlined in Sec.~\ref{sec:20220801101220} above is applied in the next section to random assemblies of circular inclusions.

\section{Numerical examples} \label{sec:20220727145908}

Note: the following numerical examples can be reproduced from the dataset  at the \texttt{Recherche Data Gouv} open data repository~\citep{bris2023}.

\medskip

In this section, we consider two applications of the EIM-based variance reduction technique to 2D random assemblies of \(N=32\) circular inhomogeneities. The size \(L\) of the square simulation box and the radius \(a\) of the disks are adjusted so as to ensure a prescribed value of the fraction \(f\) of inhomogeneities.

\begin{figure}
  \centering
  \includegraphics[width=0.5\linewidth, bb= 17.9cm 4.4cm 41cm 27.6cm]{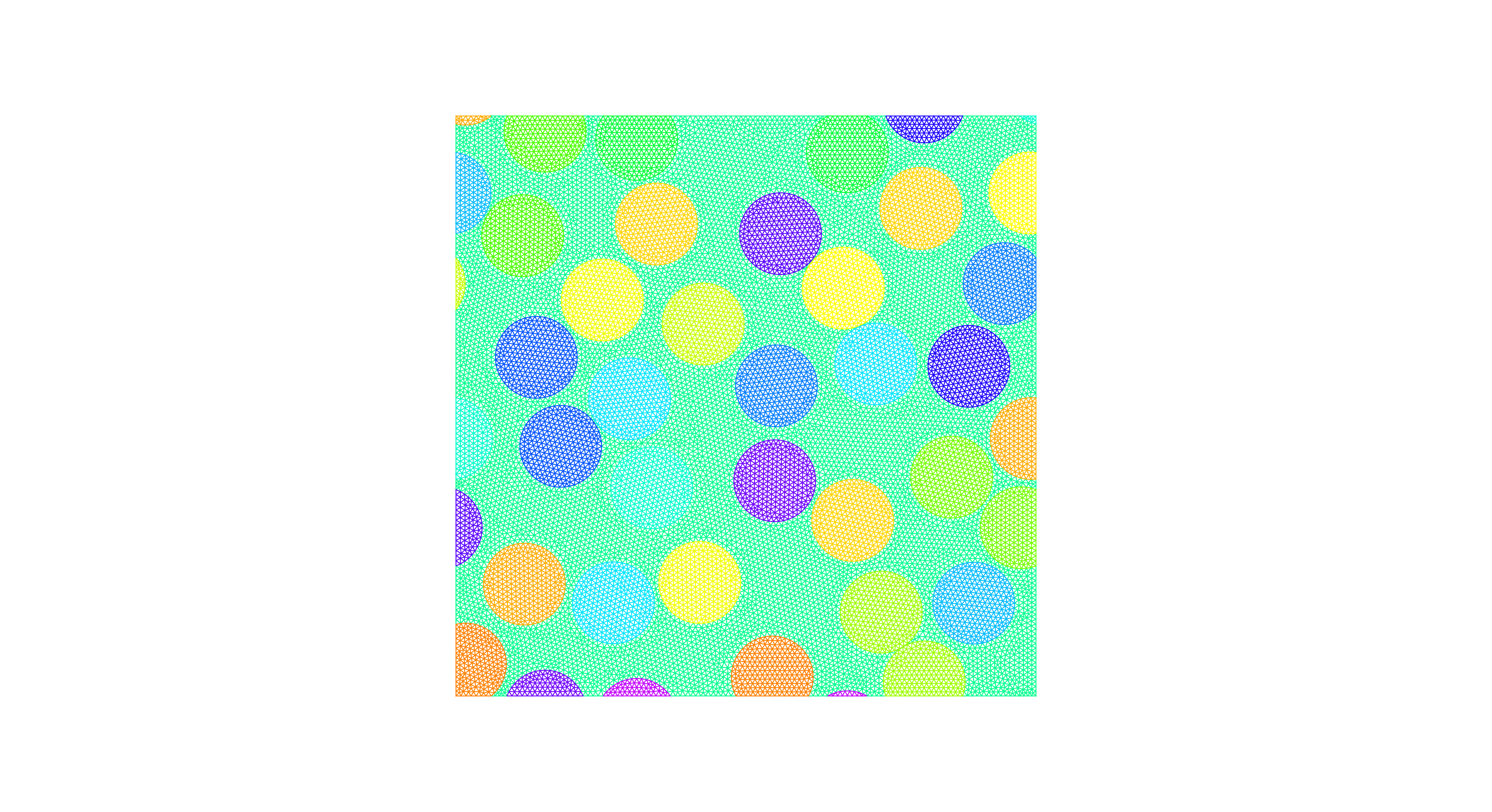}
  \caption{A realization of the microstructures considered in Sec.~\ref{sec:20220727145908}. \label{fig:20220727151638}}
\end{figure}

A standard Monte--Carlo algorithm as described by~\citet[Chap. 4]{alle1994} is implemented and used to generate all microstructures below. Starting from a centered square orthorombic lattice\footnote{\url{https://en.wikipedia.org/wiki/Orthorhombic_crystal_system\#2D_Bravais_lattices}, last retrieved 2022-07-27.}, the simulation is run for \(100\,000\) cycles (during one cycle, a trial move is proposed for each particle). The random amplitude of the trial moves is adjusted every 50 cycles in order to ensure that the acceptance ratio is about \(0.3\): if the acceptance ratio is too large (resp. too small), the amplitude of the trial moves is scaled by \(1.05\) (resp. \(0.95\)). The microstructures thus generated are subsequently meshed (mesh element size: \(h \simeq a/8\)), as shown in Fig.~\ref{fig:20220727151638}.

The (isotropic) conductivity of the matrix is \(\sigma_0\). The conductivity of the inhomogeneities is \(\chi \, \sigma_0\), where \(\chi\) denotes the contrast. We use the theoretical developments of Sec.~\ref{sec:20220801153432} and~\ref{sec:20220801101220}, where each circular inhomogeneity in $\cell$ (along with its periodic images in $\reals^d$) represents a phase. The number of realizations considered for the FEM simulations is \(M=100\). The number of realizations considered for the EIM simulations is \({\cal M} = 10\,000\).

\begin{remark}[On the periodicity of the microstructures]
  The microstructures considered until now are \emph{not necessarily periodic}. Rather, the SVE is in fact the \emph{restriction} to the finite-size unit-cell \(\cell\) of an infinite realization of the random heterogeneous material under consideration. As a consequence, the volume fraction within the unit-cell \(\cell\) is in general a random variable that may fluctuate around its expectation.

  Previous studies~\citep{lego2015,sqs2016} have shown that these fluctuations are indeed the leading sources of statistical errors and can be used as control variates (or stratification criteria) to efficiently reduce the variance. Our goal in the present article is to investigate \emph{further} sources of statistical errors, related to the spatial distribution of the phases (rather than their mere volume fractions). In order to focus on these other sources, we fix the volume fractions in the microstructures that we generate, thus killing the fluctuations of the apparent conductivity that are induced by the fluctuations of volume fractions. This is most easily achieved by \emph{enforcing \(\cell\)-periodicity of the microstructure}, which we will systematically do in the remainder of this article. In other words, all microstructures considered in the sequel are geometrically periodic. 
\end{remark}

\begin{remark}
  We again note that we are not concerned with size-effects in this work. The assemblies considered here may not be sufficiently large to provide an accurate estimate of the \emph{effective} conductivity. However, provided that the mesh is fine enough and that the number of realizations is large enough, an accurate estimate of the expectation $\ensavg[\tens{\sigma}^\app]$ of the \emph{apparent} conductivity may be obtained. We found that \(N=32\) inhomogeneities lead to interactions that are complex enough and dispersion that is large enough to assess the efficiency of the proposed variance reduction technique.
\end{remark}

\subsection{Effect of elastic contrast}

In this section, we select a high value for the fraction of inclusions, \(f = 50\%\) (this corresponds to $L / (2a) \simeq 7.08$) and investigate how the value of the contrast \(\chi\) affects the efficiency of the variance reduction. To do so, we run simulations for \(\chi \in \{10^{-6}, 10^{-5}, \ldots, 10^{-1}, 10, \ldots, 10^5, 10^6\}\).

\begin{figure}
  \centering
  \includegraphics{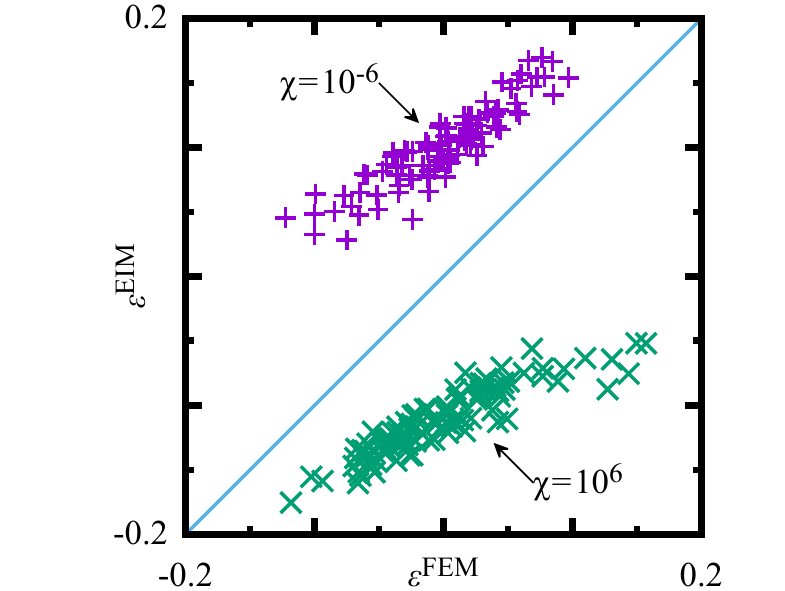}
  \caption{Plot of $\varepsilon^\EIM(\omega)$ as a function of $\varepsilon^\FEM(\omega)$ (these two quantities are defined by~\eqref{eq:20220728094325}) for $M=100$ realisations (we consider two values of the contrast: \(\chi=10^{-6}\) and \(\chi=10^6\)). \label{fig:20220728083506}}
\end{figure}

Figure~\ref{fig:20220728083506} displays, for the \(M=100\) full-field computations, the EIM estimate of \(\sigma_{11}\) vs. the FEM estimate of the same quantity. More precisely, \(M=100\) realisations of the relative error (with respect to $\ensavg(\sigma_{11}^\text{FEM})$)
\begin{equation} \label{eq:20220728094325}
  \varepsilon^\star(\omega) = \frac{\sigma_{11}^\star(\omega) - \ensavg(\sigma_{11}^\text{FEM})}{\ensavg(\sigma_{11}^\text{FEM})},
\end{equation}
where \(\star\) stands for ``FEM'' or ``EIM'', are displayed in Fig.~\ref{fig:20220728083506}. The plot shows that FEM and EIM values are strongly correlated and the proposed variance reduction method is therefore expected to perform efficiently. Note that the point clouds represented in Fig.~\ref{fig:20220728083506} are rather far from the first bisector. In other words, the EIM provides a \emph{inaccurate} estimate of the conductivity.

\medskip

\begin{figure}
  \centering
  \includegraphics{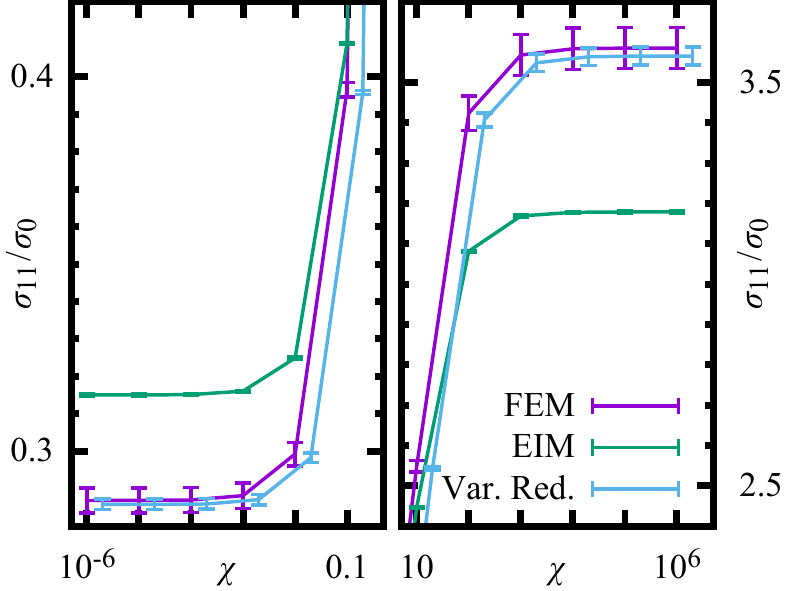}
  \caption{FEM and EIM estimates of the reduced conductivity \(\sigma_{11} / \sigma_0\), for various values of the contrast \(\chi\). For the FEM curve (resp. EIM curve), we show the empirical mean over $M$ (resp. ${\cal M}$) realisations as well as the \(99 \, \%\) confidence interval (displayed as error bars). The values obtained after variance reduction are also displayed on the plot (for the sake of clarity, this latter curve is slightly shifted to the right). \label{fig:20220728095510}}
\end{figure}

Figure~\ref{fig:20220728095510} shows the FEM and EIM estimates of the reduced conductivity \(\sigma_{11} / \sigma_0\) vs. the contrast \(\chi\) (more precisely, we plot the empirical mean over $M$ -- resp. ${\cal M}$ -- realisations as well as the \(99 \, \%\) confidence interval for the FEM estimate -- resp. the EIM estimate). As expected, \(\sigma_{11}\) increases when \(\chi\) increases. It is also observed that the error bars are much smaller for the EIM estimates than for the FEM estimates: this is simply due to the fact that the former are computed on the basis of a much larger number of realisations (${\cal M} = 10\,000 \gg M = 100$) than the latter. As already observed, the EIM does not deliver a good estimate of the conductivity (the green and purple curves of Fig.~\ref{fig:20220728095510} are significantly different). Figure~\ref{fig:20220728095510} also displays the numerical estimates of the reduced conductivity after application of the proposed variance reduction strategy. Clearly, the resulting error bars (on the blue curve) are reduced significantly in comparison to the original error bars (on the purple curve), leading to a much more accurate estimation of $\ensavg(\sigma_{11}^\text{FEM})$ at constant computational cost.

\medskip

\begin{figure}
  \centering
  \includegraphics{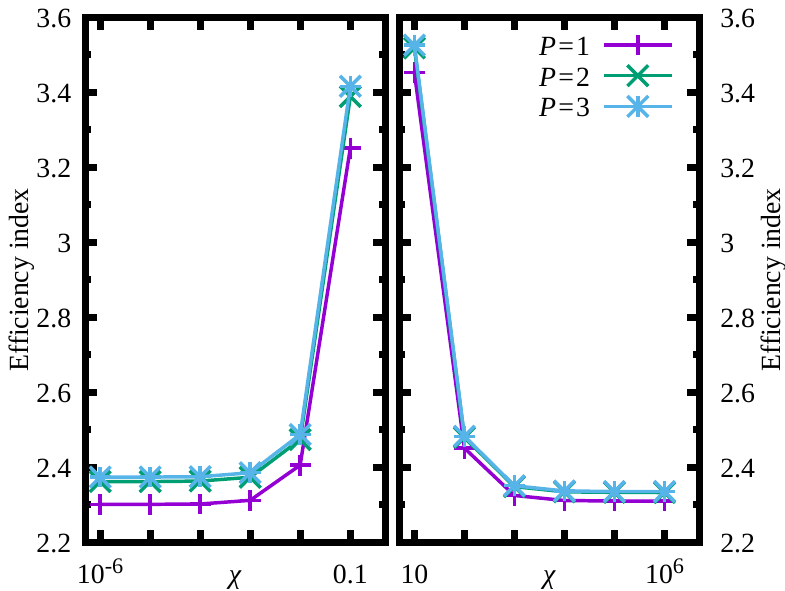}
  \caption{Efficiency index of the proposed variance reduction technique, as a function of contrast \(\chi\) and truncation parameter \(P\). \label{fig:20220728124516}}
\end{figure}

More quantitatively, Figure~\ref{fig:20220728124516} shows the efficiency index (defined at the very end of Sec.~\ref{sec:20201013092415}) of the variance reduction technique vs. the contrast \(\chi\), for $P=1$, 2 and 3. Even for highly contrasted microstructures, the standard deviation is significantly reduced by a factor of about 2.3 to 2.4. This demonstrates the value of the proposed technique. For the case considered here, the efficiency index seems to have reached a plateau for any value of $\chi \leq 10^{-4}$ (resp. any value of $\chi \geq 10^4$), which is an encouraging evidence that the efficiency may remain of the same order of magnitude for even smaller contrasts than $10^{-6}$ (resp. larger contrasts than $10^6$).

We observe that the efficiency index increases with increasing truncation parameter \(P\). However, improvements are barely noticeable for \(P=3\), and our recommended value is \(P=2\). We also observe that, for constrasts smaller than 1, the efficiency increases when $\chi$ becomes closer to 1 (and similarly for $\chi > 1$). In the limit $\chi \to 1$, the variance (denoted $\sigma_Y^2$ and $\sigma_X^2$, following the notation of Sec.~\ref{sec:20201013092415}) of the EIM and the FEM conductivities converges to 0 (when $\chi=1$, the materials is simply homogeneous), the variance $\sigma_Z^2$ of the reduced variable $Z$ converges to 0 as well, and numerical experiments show that the efficiency index $\sigma_X/\sigma_Z$ actually increases.

\begin{remark}
  When \(\chi = 10^6\) and \(P = 2\), the statistical error on \(\ensavg(\sigma_{11}^\EIM / \sigma_0)\) and on \(\ensavg(\sigma_{11}^\FEM / \sigma_0)\) (corresponding to a \(99\,\%\) confidence interval) are about \(0.0024\) and \(0.0514\), respectively. The efficiency index is about \(2.333\), and the optimum value of the variable \(\xi\) is \(\xi^\star=1.8926\). After variance reduction, the total error on \(\sigma_{11}^\app / \sigma_0\) (taking into account the uncertainty on the expectation of $\sigma_{11}^\EIM$) is
  $$
  \frac{0.0514}{2.333} + 1.8926 \times 0.0024 = 0.0266,
  $$
  leading to an \emph{effective} efficiency index of \(0.0514 / 0.0266 = 1.93\), slightly smaller than \(2.33\). Increasing ${\cal M}$ would improve this effective efficiency index.
\end{remark}

\subsection{Application to the discrimination of statistically different microstructures} \label{sec:20220801113000}

\begin{figure}
  \centering
  \includegraphics[width=0.32\linewidth, bb= 17.9cm 4.4cm 41cm 27.6cm]{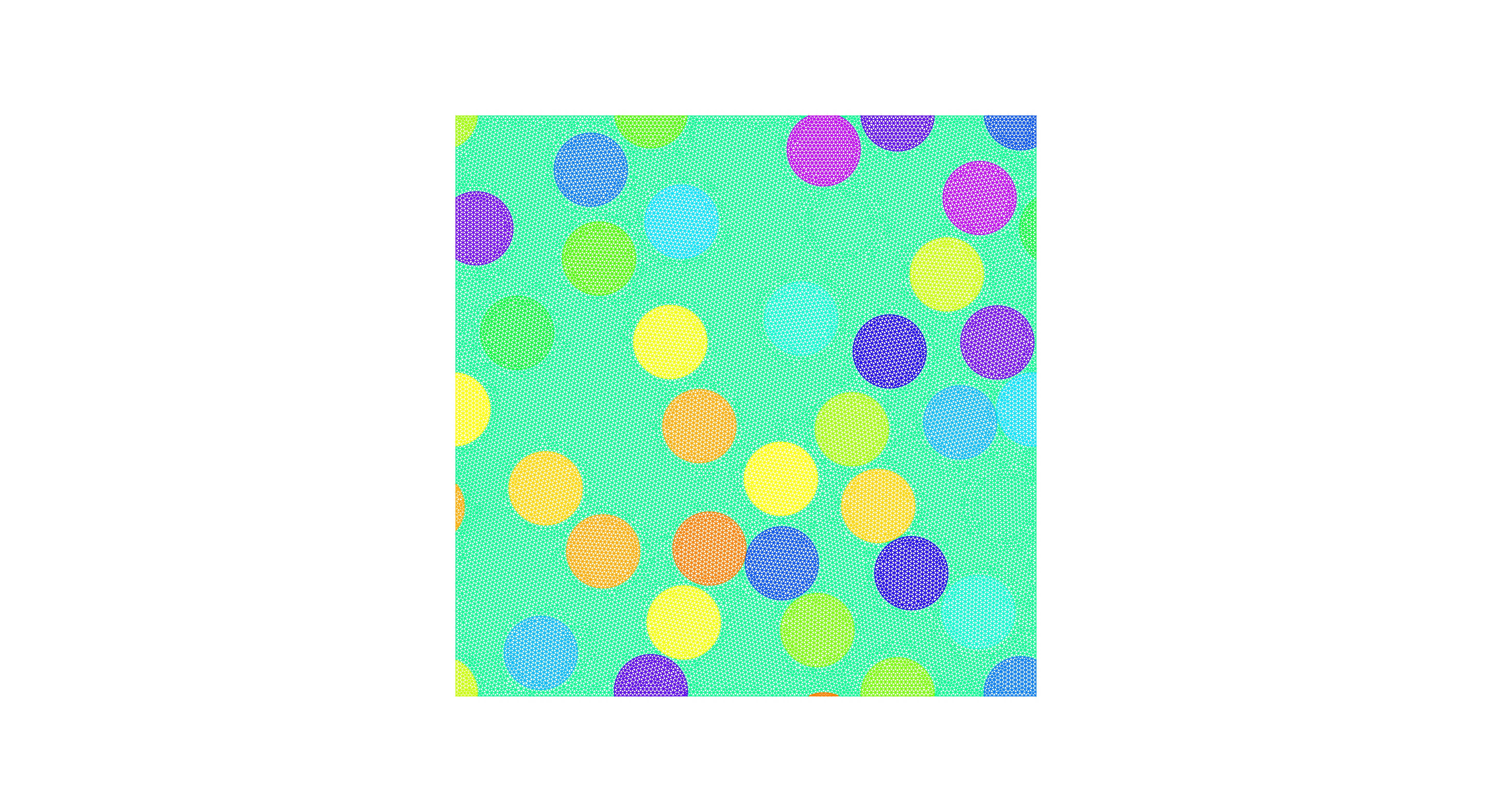}%
  \hfill
  \includegraphics[width=0.32\linewidth, bb= 17.9cm 4.4cm 41cm 27.6cm]{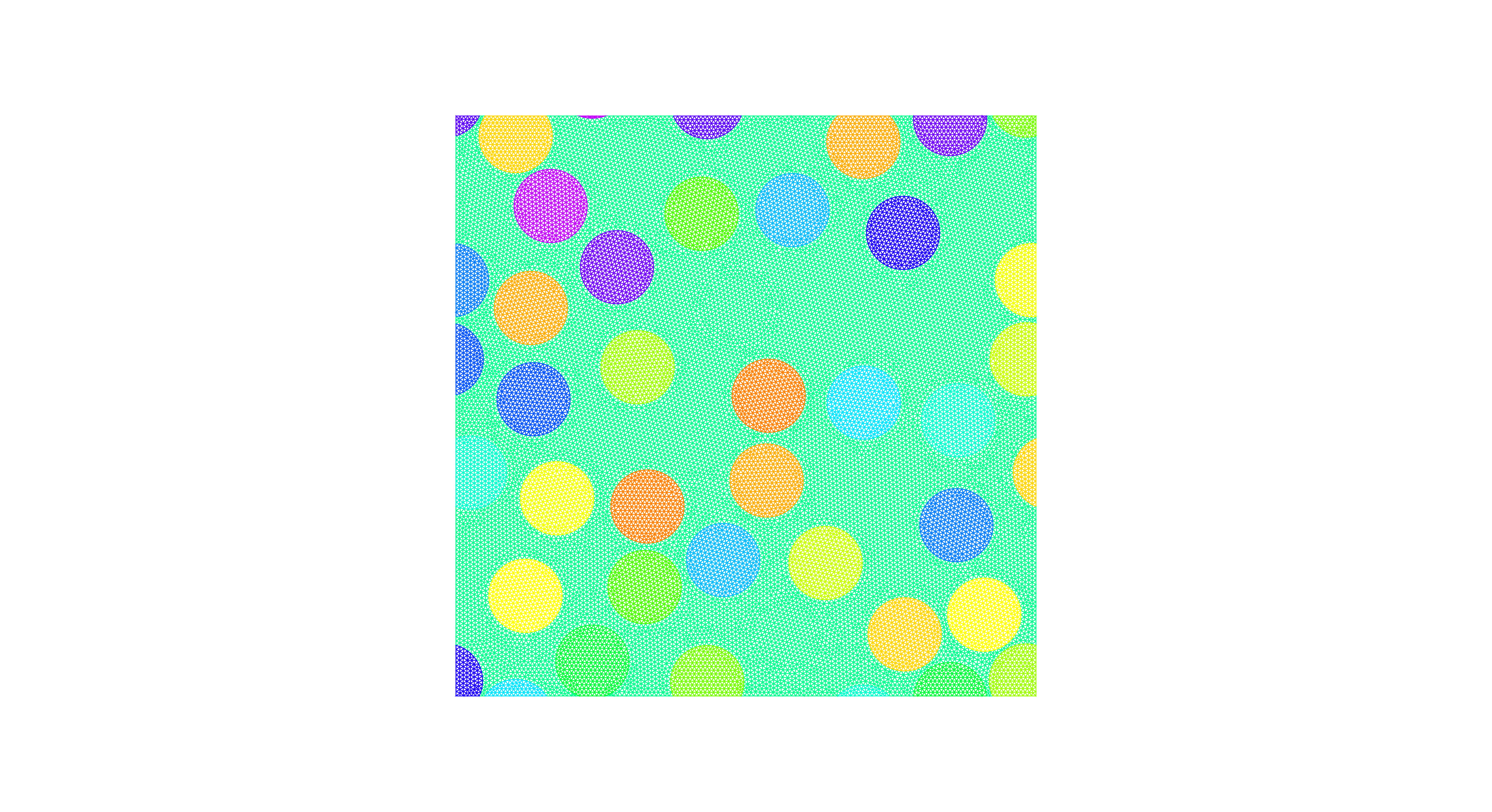}%
  \hfill
  \includegraphics[width=0.32\linewidth, bb= 17.9cm 4.4cm 41cm 27.6cm]{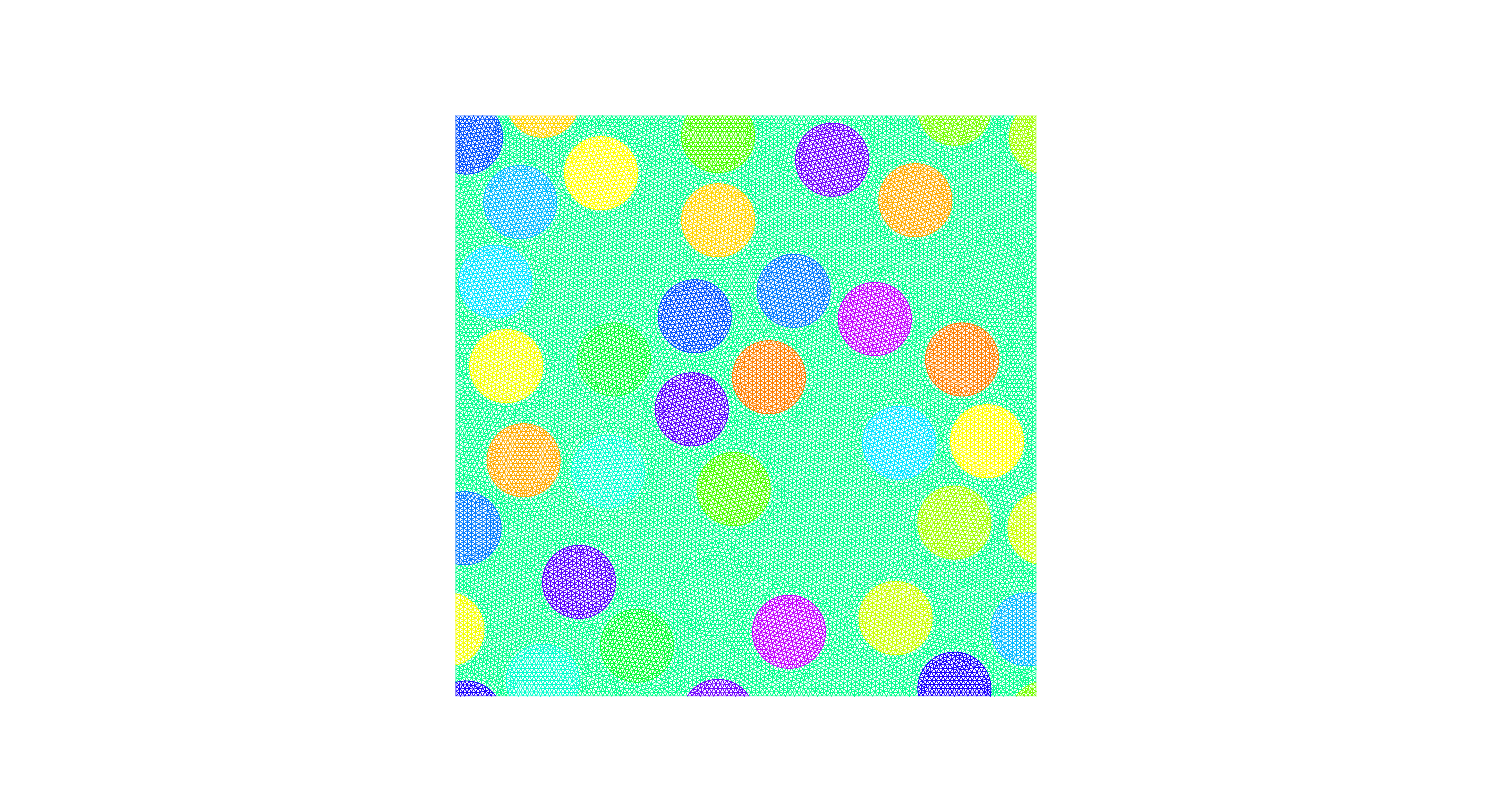}
  \caption{Examples of microstructures considered in Sec.~\ref{sec:20220801113000}. \emph{Left:} \(g / a = 0\); \emph{Center:} \(g / a = 0.121\); \emph{Right:} \(g / a = 0.236\) (we recall that \(a\) is the radius of the inhomogeneities and that \(g\) is the minimum gap between inhomogeneities). \label{fig:20220801134743}}
\end{figure}

In the present application, we consider three families of assemblies of disks. For all three families, the total number of inhomogeneities is \(N=32\). They cover \(f=40\,\%\) of the unit-cell $\cell$. As shown in Tab.~\ref{tab:20220801110058}, the three microstructures differ by the minimum allowed gap between two inhomogeneities, ranging from 0 (the inhomogeneities may touch) to 23.6 \% of their radius (see Fig.~\ref{fig:20220801134743}).

The three families of inclusions are generated with the same Monte--Carlo code as previously, starting from an initial fraction \(f'\geq f\) of inhomogeneities (we consider \(N=32\) inhomogeneities with a slightly larger radius $a'$). After equilibration, the initial radius \(a'\) of the inclusions is scaled down to the final value \(a\), so as to reach the desired volume fraction \(f\). The minimum allowed gap $g$ is defined by the relation $2a' = 2a + g$. It corresponds to the gap between two inhomogeneities of radius $a$ that would have been tangent one with each other when their radius was equal to $a'$.

\begin{table}
  \centering
  \begin{tabular}{c|c|c|c}
    \(f'\) & \(f\) & \(a / a'\) & \(g / a\) \\
    \hline
    0.40 & 0.40 & 1.000 & 0.000 \\
    0.45 & 0.40 & 0.943 & 0.121 \\
    0.50 & 0.40 & 0.894 & 0.236 \\
    \hline
  \end{tabular}
  \caption{Parameters for the Monte--Carlo simulations of Sec.~\ref{sec:20220801113000}, with the following symbols: \(f'\) and \(f\) are the initial and final inhomogeneities fraction; \(a'\) and \(a\) are the initial and final radius; \(g\) is the minimum gap between disks. \label{tab:20220801110058}}
\end{table}

\begin{figure}
  \centering
  \includegraphics{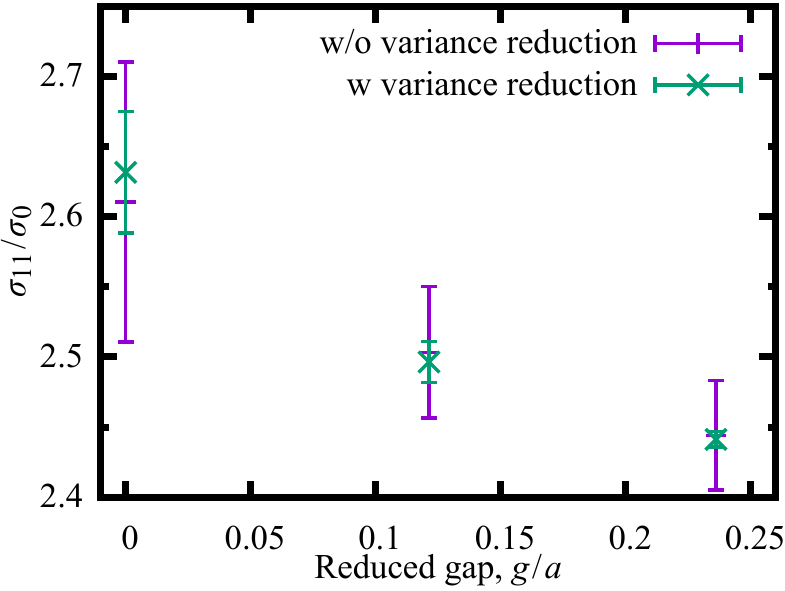}
  \caption{Estimates of the \((1,1)\) component of the apparent conductivity for the microstructures considered in Sec.~\ref{sec:20220801113000}. Without variance reduction (``w/o''), the error bars are so large that the three microstructures are statistically undistinguishable. With the variance reduction strategy (``w''), the error bars are sufficiently small so that the confidence intervals no longer overlap. \label{fig:20220801113526}}
\end{figure}

The contrast is set to \(\chi=10^6\), and the apparent conductivities of the microstructures are estimated with \(M=10\) finite element simulations (\(M\) is kept very low on purpose in this application). The results are shown in Fig.~\ref{fig:20220801113526}, where it is observed that the error bars are too large to allow for the discrimination of the three types of microstructures. The EIM-based variance reduction strategy is next applied (with \({\cal M} = 10\,000\)). As can be seen in Fig.~\ref{fig:20220801113526}, discrimination is possible after variance reduction: the apparent conductivity \emph{does} depend on the gap, and actually decreases when the gap increases.

\begin{remark}
  In the present article, no attempt has been made to optimize the number ${\cal M}$ of realizations used to evaluate the ensemble average of the surrogate model. The large value (${\cal M} = 10\,000$) selected here leads to a very small statistical error on the evaluation of $\ensavg(\sigma^\EIM)$. We have checked that, although the error bars on the quantity of interest get larger, reducing ${\cal M}$ down to ${\cal M} = 100$ still allows to separate the three families of microstructures (see Table~\ref{tab:20220801143525}).
\end{remark}

\begin{table}
  \centering
  \begin{tabular}{c|c|c}
    \(f'\) & Lower bound & Upper bound \\
    \hline
    0.40 & 2.563 & 2.701 \\
    0.45 & 2.467 & 2.525 \\
    0.50 & 2.425 & 2.426 \\
    \hline
  \end{tabular}
  \caption{Bounds of the \(99\,\%\) confidence interval on the \((1,1)\) component of the apparent conductivity for the three microstructures considered in Sec.~\ref{sec:20220801113000}, in the case when ${\cal M} = 100$. The values shown in this table account for the statistical errors on the surrogate model. \label{tab:20220801143525}}
\end{table}

\section{Conclusion} \label{sec:conc}

Within the framework of linear conductivity, we have proposed a variance reduction strategy based on the use of a surrogate model, here the equivalent inclusion method, for the estimation of the macroscopic (apparent) properties. This strategy has been implemented in a two-dimensional setting, for assemblies of disks. Even for high values of the matrix-inhomogeneity conductivity contrast (namely $10^6$ and $10^{-6}$), the proposed strategy performs very satisfactorily: its efficiency index (which quantifies the accuracy gain at constant computational cost) remains above 2.2. This is quite remarkable, since the equivalent inclusion method is known to deliver poor estimates of the apparent properties (especially in the regime of high contrast and/or high volume fraction). Still, the EIM method seems to be able to capture some salient features of the microstructures that lead to a high correlation with the true apparent conductivity.

There are several perspectives to the present work. First, the framework presented here readily extends to linear elasticity: we refer to~\citet{bris2014} for a presentation of the equivalent inclusion method in that setting.

Second, owing to the similarities between the equivalent inclusion method and the \emph{self-consistent clustering} method~\citep{schn2019b}, a similar variance reduction strategy could be proposed in situations where the microstructure is not an assembly of elementary geometrical objects. A full-field approach based on the fast Fourier transform~\citep{moul1994,moul1998} must then be adopted. Of course, this increases the cost of the reduced model (one evaluation of the surrogate model being roughly equivalent to a few iterations of the full-field simulation), but the strategy might still be competitive. Likewise, there are deep connexions between the equivalent inclusion method and the recent works by~\citet{to2016,nguy2016a,to2017}, that should be reassessed from the point of view of variance reduction.

Finally, a heuristic correction has been proposed to a recent formula by~\citet{zece2021}. This formula (actually its corrected version) is used here to efficiently compute the influence tensors when periodic boundary conditions are considered. We have numerically investigated the validity of this corrected formula in the case of circular inclusions embedded in a square unit-cell. Extending the correction to other settings and producing a mathematical proof of the corrected formula is another exciting perspective.

\appendix
\renewcommand{\thesection}{\Alph{section}}

\section{$H(\div,\cell)$ space} \label{app:H_div}

We recall here some results about the $H(\div,\cell)$ space. For any bounded domain $\cell$ of $\reals^d$, we recall that
$$
H(\div,\cell):= \left\{ v \in (L^2(\cell))^d, \ \ \div v \in L^2(\cell) \right\},
$$
and that $H(\div,\cell)$ is a Hilbert space. The space $\left(\mathcal{C}^\infty(\overline{\cell})\right)^d$ is dense in $H(\div,\cell)$. Consider the normal trace application
$$
\gamma : \left\{
\begin{array}{ccc}
\left(\mathcal{C}^\infty(\overline{\cell})\right)^d & \to & \mathcal{C}^0(\partial \cell)\\
v & \mapsto & (v \cdot n)|_{\partial \cell}
\end{array}
\right. ,
$$
where $n$ denotes the unit exterior normal vector to $\partial \cell$. The application $\gamma$ can be uniquely extended (as a continuous application) on the whole space $H(\div,\cell)$, which in particular implies that the normal trace $v \cdot n$ of any field $v \in H(\div,\cell)$ is well-defined.

\section{Fourier series and Fourier transforms} \label{sec:20201006143813}

We define here our conventions for the Fourier transform of a periodic function. We consider a \(\cell\)-periodic tensor field \(\tens{T}\), with components that are square-integrable over \(\cell\) (we recall that \(\cell = (-L/2,L/2)^d\)). For any $\vec n \in \integers^d$, its Fourier coefficient \(\widetilde{\tens{T}}_{\vec n}\) is defined by
\begin{equation*} 
  \widetilde{\tens{T}}_{\vec n} = \frac{1}{| \cell |} \int_\cell \tens{T}(\vec x) \, \E^{-\I \, \vec k_{\vec n} \cdot \vec x} \, \D \vec x,
  \quad \text{with} \quad
  \vec k_{\vec n} = \frac{2\pi \, \vec n}{L}.
\end{equation*}
The synthesis formula then reads
\begin{equation*} 
  \tens{T}(\vec x) = \sum_{\vec n \in \integers^d} \widetilde{\tens{T}}_{\vec n} \, \E^{\I \, \vec k_{\vec n} \cdot \vec x},
\end{equation*}
where the above right-hand side is a converging series in $(L^2(\cell))^{d \times d}$.


\section{Variational formulation of the equivalent inclusion method} \label{sec:20201013074424}

Multiplying~\eqref{eq:20200616063758} with an arbitrary test function \(\vec Q\) (a vector-valued field with square integrable components) and averaging over the unit-cell delivers the weak form of the Lippmann--Schwinger equation~\eqref{eq:20200616063758}, that is written as Problem~\ref{pb:20201012134607} below:

\begin{problem} \label{pb:20201012134607}
  Find \(\vec P \in (L^2(\cell))^d\) such that, for any \(\vec Q \in (L^2(\cell))^d\),
  \begin{equation} \label{eq:20200617061935}
    \langle \vec Q \cdot \bigl(\tens{\sigma}-\tens{\sigma}_0\bigr)^{-1} \, \vec P \rangle + \langle \vec Q \cdot \tens{\Gamma}_0(\vec P) \rangle = \overline{\vec E} \cdot \langle \vec Q \rangle.
  \end{equation}
\end{problem}
Note that the bilinear form in the above variational equation is symmetric (the Green operator \(\tens{\Gamma}_0\) owes its symmetry to the symmetry of the conductivity tensor \(\tens{\sigma}_0\)). The well-posedness of~\eqref{eq:20200617061935} can be studied directly (without using the equivalence of the Lippmann--Schwinger equation~\eqref{eq:20200616063758} with Problem~\ref{pb:20201012143332}). Such a study (which has been been performed in, e.g.,~\citet{brisard_dormieux12,schneider15}) opens the way to establishing the well-posedness of Galerkin discretizations of~\eqref{eq:20200617061935}.

The variational form of the EIM results from the Galerkin discretization of~\eqref{eq:20200617061935} with phase-wise constant trial and test fields:
\begin{equation} \label{eq:20201012135048}
  \vec P(\vec x) = \sum_{\alpha=1}^N \chi_\alpha(\vec x) \, \vec P_\alpha \quad \text{and} \quad \vec Q(\vec x) = \sum_{\alpha=1}^N \chi_\alpha(\vec x) \, \vec Q_\alpha,
\end{equation}
where \(\vec P_\alpha, \vec Q_\alpha\in\reals^d\) are constant vectors (we recall that $\chi_\alpha$ is the indicator function of the phase $\alpha$). Inserting~\eqref{eq:20201012135048} in~\eqref{eq:20200617061935} and using the definition~\eqref{eq:20200721095808} of the influence tensors $\tens{\Gamma}_{\alpha\beta}$ yields the discrete variational Problem~\ref{pb:20201012134510} below.

\begin{problem} \label{pb:20201012134510}
  Find \(\vec P_1, \dots, \vec P_N\in\reals^d\) such that, for all \(\vec Q_1, \ldots, \vec Q_N\in\reals^d\),
  \begin{multline*}
    \sum_{\alpha=1}^N f_\alpha \, \vec Q_\alpha \cdot \bigl(\tens{\sigma}_\alpha-\tens{\sigma}_0\bigr)^{-1} \, \vec P_\alpha
    \\
    + \sum_{\alpha,\beta=1}^N f_\alpha \, \vec Q_\alpha \cdot \tens{\Gamma}_{\alpha\beta} \, \vec P_\beta = \sum_{\alpha=1}^N f_\alpha \, \overline{\vec E} \cdot \vec Q_\alpha.
  \end{multline*}
\end{problem}
Since the vectors \(\vec Q_1\), \ldots, \(\vec Q_N\) are arbitrary, Problem~\ref{pb:20201012134510} results in the linear system~\eqref{eq:20200629090809}.

\section{On the influence tensor (periodic BCs)} \label{sec:20201006153018}

We derive here the expression~\eqref{eq:20220705185558} of the influence tensor of two inhomogeneities in a periodic setting. Recalling that the influence tensor $\tens{\Gamma}_{\alpha\beta}$ is defined by~\eqref{eq:20200721095808}, we first note, owing to \(\tens{\Gamma}_0\) being self-adjoint, that \(f_\alpha \, \tens{\Gamma}_{\alpha\beta} = f_\beta \, \tens{\Gamma}_{\beta\alpha}\), where we recall that \(f_\alpha\) denotes the volume fraction of inhomogeneity \(\alpha\).

When the phase \(\beta\) is subjected to the constant polarization \(\vec P_\beta\), we have \(\vec P(\vec x) = \chi_\beta(\vec x) \, \vec P_\beta\) at any point \(\vec x \in \cell\). In view of the expression~\eqref{eq:20201006145808} of the Green operator \(\tens{\Gamma}_0\) in the Fourier space, we can express the induced electric field $\vec E = -\tens{\Gamma}_0(\chi_\beta \, \vec P_\beta)$ at any point as
$$
\vec E(\vec x) = -\sum_{\vec n \in \integers^d} \widetilde{\chi}_{\beta, \vec n} \, \exp\bigl(\I \, \vec k_{\vec n} \cdot \vec x\bigr) \, \widehat{\tens{\Gamma}}_0(\vec k_{\vec n}) \, \vec P_\beta,
$$
where the Fourier coefficients \(\widetilde{\chi}_{\beta, \vec n}\) of the indicator function $\chi_\beta$ are defined by~\eqref{eq:20220705190239}.

The above expression is then averaged over the phase \(\alpha\). In view of~\eqref{eq:20200721095808}, we have that, for any \(\vec P_\beta\),
$$
\tens{\Gamma}_{\alpha\beta} \, \vec P_\beta = -| \Omega_\alpha \cap \cell |^{-1} \int_\cell \chi_\alpha(\vec x) \, \vec E(\vec x) \, \D \vec x,
$$
from which we deduce, observing that \(| \Omega_\alpha \cap \cell | = f_\alpha \, | \cell |\), that
$$
f_\alpha \, \tens{\Gamma}_{\alpha\beta} = | \cell |^{-1} \int_\cell \sum_{\vec n \in \integers^d} \chi_\alpha(\vec x) \, \widetilde{\chi}_{\beta, \vec n} \, \widehat{\tens{\Gamma}}_0(\vec k_{\vec n}) \exp \bigl( \I \, \vec k_{\vec n} \cdot \vec x \bigr) \, \D \vec x.
$$
Recognizing the expression~\eqref{eq:20220705190239} of \(\widetilde{\chi}_{\alpha, \vec n}\), Equation~\eqref{eq:20220705185558} is eventually retrieved.

\medskip

We next derive a closed form expression of the \emph{trace} of the influence tensor \(\tens{\Gamma}_{\alpha\beta}\), in the case of isotropic reference materials. From the general expression~\eqref{eq:20220705185558}, we infer
$$
f_\alpha \, \tr\tens{\Gamma}_{\alpha\beta} = \sum_{\vec n \in \integers^d} \widetilde{\chi}_{\alpha, \vec n}^\ast \ \widetilde{\chi}_{\beta, \vec n} \ \tr\left[ \widehat{\tens{\Gamma}}_0(\vec k_{\vec n}) \right].
$$
In view of~\eqref{eq:20200722092039}, we have $\tr\Bigl[ \widehat{\tens{\Gamma}}_0(\vec 0) \Bigr] = 0$ and $\tr\Bigl[ \widehat{\tens{\Gamma}}_0(\vec k) \Bigr] = \sigma_0^{-1}$ for any $\vec k \neq \vec 0$. We therefore deduce that
$$
\sigma_0 \, f_\alpha \, \tr\tens{\Gamma}_{\alpha\beta} = \sum_{\vec n \in \integers^d} \widetilde{\chi}_{\alpha, \vec n}^\ast \ \widetilde{\chi}_{\beta, \vec n} - \widetilde{\chi}_{\alpha, \vec 0}^\ast \ \widetilde{\chi}_{\beta, \vec 0}.
$$
Parseval's theorem is applied to the first term, and we notice that $\widetilde{\chi}_{\alpha, \vec 0} = \langle \chi_\alpha \rangle = f_\alpha$ in the second term. We thus obtain
$$
\sigma_0 \, f_\alpha \, \tr\tens{\Gamma}_{\alpha\beta} = \langle \chi_\alpha \, \chi_\beta \rangle - f_\alpha \, f_\beta.
$$
If \(\alpha \neq \beta\), the domains \(\Omega_\alpha\) and \(\Omega_\beta\) do not overlap, and \(\langle \chi_\alpha \, \chi_\beta \rangle = 0\). If \(\alpha = \beta\), \(\chi_\alpha\,\chi_\beta = \chi_\alpha\) and \(\langle \chi_\alpha \, \chi_\beta \rangle = f_\alpha\). We eventually obtain
\begin{equation} \label{eq:20220707143902}
  \tr\tens{\Gamma}_{\alpha\beta} =
  \begin{cases}
    -f_\beta / \sigma_0 & \text{when $\alpha \neq \beta$}, \\
    \bigl(1 - f_\alpha\bigr) / \sigma_0 & \text{when $\alpha = \beta$}.
  \end{cases}
\end{equation}

\medskip

To close this section, we consider the particular case when $d=2$ and when the phase $\alpha$ occupies a disk \(\Omega_\alpha\), centered at \(\vec x_\alpha\) and of radius \(a_\alpha\). We then have
\begin{align*}
  \widetilde{\chi}_{\alpha, \vec n}
  & = | \cell |^{-1} \int_{| \vec x - \vec x_\alpha | \leq a_\alpha} \exp\bigl(-\I \, \vec k_{\vec n} \cdot \vec x\bigr) \, \D \vec x \\
  & = | \cell |^{-1} \int_{| \vec r | \leq a_\alpha} \exp\bigl(-\I \, \vec k_{\vec n} \cdot \bigl(\vec r + \vec x_\alpha \bigr)\bigr) \, \D \vec r \\
  & = | \cell |^{-1} \, \E^{-\I \, \vec k_{\vec n} \cdot \vec x_\alpha} \int_{| \vec r | \leq a_\alpha} \exp\bigl(-\I \, \vec k_{\vec n} \cdot \vec r \bigr) \, \D \vec r \\
  & = | \cell |^{-1} \, \E^{-\I \, \vec k_{\vec n} \cdot \vec x_\alpha} \int_0^{a_\alpha} \int_0^{2\pi} \exp\bigl(-\I \, k_{\vec n} \, r \cos\theta \bigr) \, r \, \D\theta \, \D r \\
  & = 2\pi \, | \cell |^{-1} \, \E^{-\I \, \vec k_{\vec n} \cdot \vec x_\alpha} \int_0^{a_\alpha} \besselj_0(k_{\vec n} \, r) \, r \, \D r \\
  & = 2\pi \, k_{\vec n}^{-2} \, | \cell |^{-1} \, \E^{-\I \, \vec k_{\vec n} \cdot \vec x_\alpha} \int_0^{k_{\vec n} \, a_\alpha} \besselj_0(\xi) \, \xi \, \D \xi\\
  & = 2\pi \, a_\alpha \, k_{\vec n}^{-1} \, | \cell |^{-1} \, \E^{-\I \, \vec k_{\vec n} \cdot \vec x_\alpha} \, \besselj_1(k_{\vec n} \, a_\alpha),
\end{align*}
where we recall that $k_{\vec n} = | \vec k_{\vec n} |$. In the above derivation, $\besselj_0$ is the Bessel function of the first kind defined by
$$
\besselj_0(r) = \frac{1}{2\pi} \int_{-\pi}^{\pi} \exp\bigl(-\I \, r \sin\theta \bigr) \, \D\theta,
$$
and $\besselj_1$ is the Bessel function of the first kind defined by
$$
\besselj_1(r) = \frac{1}{2\pi} \int_{-\pi}^{\pi} \exp\bigl(\I \, (\theta - r \sin\theta) \bigr) \, \D\theta.
$$
Note that, in the above derivation, we have used Eqs.~10.9.1 and~10.6.6 from the \emph{Digital Library of Mathematical Functions}~\citep{dlmf2022}. We have thus retrieved the expression~\eqref{eq:20220706162136}.

\section{On the influence tensor (BCs at infinity)} \label{sec:20201007151753}

In the present appendix, we restrict ourselves to the two-dimensional setting and we derive the closed form expression of the influence tensor \(\tens{\Gamma}_{12}^\infty\) (defined in Sec.~\ref{sec:20220729143106}, see in particular~\eqref{eq:def_Gamma_infini}) of two circular inclusions \(\Omega_1\) and \(\Omega_2\) with radii \(a_1\) and \(a_2\), centered at \(\vec x_1\) and \(\vec x_2 = \vec x_1 + \vec r_{12}\) (see Fig.~\ref{fig:20220729171357}). We also derive the expression of the self-influence tensor \(\tens{\Gamma}_{11}^\infty\). The two inclusions \(\Omega_1\) and \(\Omega_2\) are embedded in a homogeneous, isotropic, infinite matrix with conductivity \(\tens{\sigma}_0 = \sigma_0 \, \tens{I}\). We also assume that the two inclusions do not overlap: $| \vec r_{12} | \geq a_1+a_2$.

\newlength{\sbxB}
\newlength{\sbyB}
\newlength{\sbrB}

\setlength{\sbxA}{-1.8cm}
\setlength{\sbyA}{-5mm}
\setlength{\sbrA}{1cm}

\setlength{\sbxB}{3.2cm}
\setlength{\sbyB}{-5mm}
\setlength{\sbrB}{1.5cm}

\begin{figure}
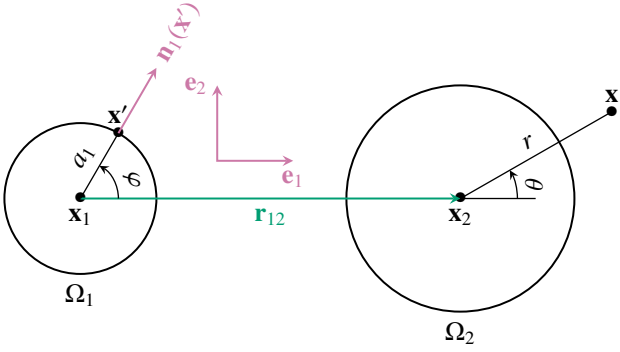

  \centering
  \begin{pgfpicture}
    \pgfslopedattimetrue
    \pgfsetlinewidth{\pgfcadnormal}
    \pgfcircle{\pgfpoint{\sbxA}{\sbyA}}{\sbrA}
    \pgfcircle{\pgfpoint{\sbxB}{\sbyB}}{\sbrB}
    \pgfusepath{stroke}

    \begin{pgfscope}
      \pgftransformshift{\pgfpoint{\sbxA}{\sbyA}}
      \pgftext{\(\bullet\)}
      \pgfnode{rectangle}{north}{\(\vec x_1\)}{}{\pgfusepath{}}
      \pgftransformshift{\pgfpoint{0pt}{-\sbrA}}
      \pgfnode{rectangle}{north}{\(\Omega_1\)}{}{\pgfusepath{}}
    \end{pgfscope}
    \begin{pgfscope}
      \pgftransformshift{\pgfpoint{\sbxB}{\sbyB}}
      \pgftext{\(\bullet\)}
      \pgfnode{rectangle}{north}{\(\vec x_2\)}{}{\pgfusepath{}}
      \pgftransformshift{\pgfpoint{0pt}{-\sbrB}}
      \pgfnode{rectangle}{north}{\(\Omega_2\)}{}{\pgfusepath{}}
    \end{pgfscope}

    \begin{pgfscope}
      \pgfsetarrowsend{stealth}
      \pgfsetstrokecolor{sbgreen}
      \pgfcaddecoratedline{0.5}{\pgfpoint{\sbxA}{\sbyA}}{\pgfpoint{\sbxB}{\sbyB}}{north}{\(\color{sbgreen}\vec r_{12}\)}
    \end{pgfscope}
    \begin{pgfscope}
      \color{sbpurple}
      \pgfsetstrokecolor{sbpurple}
      \pgfsetfillcolor{sbpurple}
      \pgfsetarrows{stealth-stealth}
      \pgfpathmoveto{\pgfpoint{0}{\sbunitvectorlength}}
      \pgfpathlineto{\pgfpointorigin}
      \pgfpathlineto{\pgfpoint{\sbunitvectorlength}{0}}
      \pgfusepath{stroke}
      \begin{pgfscope}
        \pgftransformshift{\pgfpoint{\sbunitvectorlength}{0}}
        \pgfnode{rectangle}{north}{\(\vec e_1\)}{}{\pgfusepath{}}
      \end{pgfscope}
      \begin{pgfscope}
        \pgftransformshift{\pgfpoint{0}{\sbunitvectorlength}}
        \pgfnode{rectangle}{east}{\(\vec e_2\)}{}{\pgfusepath{}}
      \end{pgfscope}
    \end{pgfscope}

    \begin{pgfscope}
      \pgftransformshift{\pgfpoint{\sbxA}{\sbyA}}
      \begin{pgfscope}
        \pgftransformshift{\pgfpointpolar{60}{\sbrA}}
        \pgfcoordinate{sbM}{\pgfpointorigin}
        \pgftext{\(\bullet\)}
        \pgfnode{rectangle}{south}{\(\vec x'\)}{}{\pgfusepath{}}
      \end{pgfscope}
      \pgfsetlinewidth{\pgfcadthin}
      \pgfcaddecoratedline{0.5}{\pgfpointorigin}{\pgfpointanchor{sbM}{center}}{south}{\(a_1\)}
      \pgfusepath{stroke}

      \pgfcadmarkangle{60}{0.5\sbrA}{south}{\(\varphi\)}

      \pgfsetlinewidth{\pgfcadnormal}
      \pgfsetarrowsend{stealth}
      \pgfsetstrokecolor{sbpurple}
      \pgftransformshift{\pgfpointanchor{sbM}{center}}
      \pgftransformrotate{60}
      \pgfpathmoveto{\pgfpointorigin}
      \pgfpathlineto{\pgfpoint{\sbunitvectorlength}{0}}
      \pgfusepath{stroke}
      \pgftransformshift{\pgfpoint{\sbunitvectorlength}{0}}
      \pgfnode{rectangle}{west}{\(\color{sbpurple}\vec n_1(\vec x')\)}{}{\pgfusepath{}}
    \end{pgfscope}

    \begin{pgfscope}
      \pgftransformshift{\pgfpoint{\sbxB}{\sbyB}}
      \begin{pgfscope}
        \pgftransformshift{\pgfpointpolar{30}{2.3cm}}
        \pgfcoordinate{sbP}{\pgfpointorigin}
        \pgftext{\(\bullet\)}
        \pgfnode{rectangle}{south}{\(\vec x\)}{}{\pgfusepath{}}
      \end{pgfscope}
      \pgfsetlinewidth{\pgfcadthin}
      \pgfcaddecoratedline{0.5}{\pgfpointorigin}{\pgfpointanchor{sbP}{center}}{south}{\(r\)}
      \pgfpathmoveto{\pgfpointorigin}
      \pgfpathlineto{\pgfpoint{.66\sbrB}{0}}
      \pgfusepath{stroke}

      \pgfcadmarkangle{30}{0.5\sbrB}{south}{\(\theta\)}
    \end{pgfscope}

  \end{pgfpicture}
  \caption{Geometric setting for the derivation of \(\tens{\Gamma}_{12}^\infty\). \label{fig:20220729171357}}
\end{figure}

\subsection{Analytical solution}

Inclusion \(\Omega_2\) is subjected to the (constant) polarization \(\vec P\). We seek the (average) induced electric field in \(\Omega_1\). The problem at hand is therefore (see~\eqref{eq:edp_espace_infini1}--\eqref{eq:edp_espace_infini4})
\begin{gather}
  \label{eq:20211123105553}
  \div \vec J = 0 \quad \text{in $\reals^2$}, \\
  \vec J = \tens{\sigma}_0 \vec E + \vec P \quad \text{in $\Omega_2$}, \\
  \vec J = \tens{\sigma}_0 \vec E \quad \text{in $\reals^2 \setminus \Omega_2$}, \\
  \label{eq:20211123105612}
  \vec E = -\grad\Phi \quad \text{in $\reals^2$}.
\end{gather}
The boundary conditions are encoded in the fact that the scalar field \(\Phi\) (which belongs to $L^2_{\rm loc}(\reals^2)$) is such that \(\vec E\) (and therefore \(\vec J\)) has components that are square integrable over \(\reals^2\): $\vec E \in (L^2(\reals^2))^2$. Introducing the center-to-center distance \(r_{12} = | \vec r_{12} |\) and the unit vectors \(\vec e_1 = \vec r_{12} / r_{12}\) and \(\vec e_2\) such that \(\angle(\vec e_1, \vec e_2) = \pi / 2\) (see Fig.~\ref{fig:20220729171357}), the solution to problem~\eqref{eq:20211123105553}--\eqref{eq:20211123105612} is classically found in cylindrical coordinates \(\bigl(r, \theta\bigr)\) with pole \(\vec x_2\) and polar axis \(\vec e_1\):
$$
\vec x  =\vec x_2 + r \cos\theta \, \vec e_1 + r \sin\theta \, \vec e_2.
$$
The (unique up to the addition of a constant) electric potential is seeked in the form
$$
\Phi(\vec x) = f(r) \, \cos\theta + g(r) \, \sin\theta,
$$
and it can be shown that \(f\) and \(g\) both satisfy the same differential equation,
$$
r^2 f'' + r f' - f = 0 \quad \text{and} \quad r^2 g'' + r g' - g = 0,
$$
for $r \in (0,a_2)$ and for $r \in (a_2,+\infty)$. This implies that, on each of the two intervals $(0,a_2)$ and $(a_2,+\infty)$, \(f\) and \(g\) are linear combinations of \(r\) and \(1/r\). Since $\Phi$ should be square integrable in the neighboorhood of the origin, and since $\nabla \Phi$ should belong to $(L^2(\reals^2 \setminus \Omega_2))^2$, we obtain that
$$
f(r) = C_{f,{\rm int}} \, \frac{r}{a_2} \quad \text{on $(0,a_2)$}, \qquad f(r) = C_{f,{\rm ext}} \, \frac{a_2}{r} \quad \text{on $(a_2,+\infty)$},
$$
and likewise for $g$. The integration constants $C_{f,{\rm int}}$, $C_{f,{\rm ext}}$, $C_{g,{\rm int}}$ and $C_{g,{\rm ext}}$ are found from the continuity of the electric potential (which implies that $C_{f,{\rm int}} = C_{f,{\rm ext}}$ and $C_{g,{\rm int}} = C_{g,{\rm ext}}$) and of the radial flux, leading to
\begin{equation} \label{eq:20201006160008}
  \Phi(\vec x)=
  \begin{cases}
    \dfrac{1}{2\sigma_0} \, \vec P \cdot \bigl(\vec x-\vec x_2\bigr) & \text{if $|\vec x-\vec x_2 | \leq a_2$}, \\ \noalign{\vskip 3pt}
    \dfrac{a_2^2}{2\sigma_0} \, \dfrac{\vec P \cdot \bigl(\vec x-\vec x_2\bigr)}{|\vec x-\vec x_2 |^2} & \text{otherwise.}
  \end{cases}
\end{equation}
This expression is next used to evaluate the average electric field over the two inclusions.

\subsection{Evaluation of the influence tensor \(\tens{\Gamma}_{12}^\infty\)}

By definition, $-\tens{\Gamma}_{12}^\infty \, \vec P$ is the average of the electric field \(\vec E = -\grad\Phi\) over the inclusion \(\Omega_1\). Since the inclusions do not overlap, the electric potential in inclusion \(\Omega_1\) reads (see~\eqref{eq:20201006160008})
\begin{equation} \label{eq:20201007145154}
  \Phi(\vec x) = \frac{a_2^2}{2\sigma_0} \, \frac{\vec P \cdot \bigl(\vec x-\vec x_2\bigr)}{| \vec x-\vec x_2 |^2}.
\end{equation}
We use Green's formula to transform the average over $\Omega_1$ into an integral over the boundary of the inclusion:
\begin{equation} \label{eq:20201006160342}
  \pi \, a_1^2 \ \tens{\Gamma}_{12}^\infty \, \vec P
  = \int_{\Omega_1} \grad\Phi
  = \int_{\partial \Omega_1} \Phi(\vec x') \, \vec n_1(\vec x') \, \D s,
\end{equation}
where \(\vec n_1\) denotes the outer unit normal to \(\partial \Omega_1\) and \(s\) is the arc-length measured on the boundary \(\partial \Omega_1\). Inserting~\eqref{eq:20201007145154} in~\eqref{eq:20201006160342}, we find
$$
\tens{\Gamma}_{12}^\infty = \frac{1}{2\pi\sigma_0} \, \frac{a_2^2}{a_1^2} \, \int_{\vec x' \in \partial \Omega_1} \vec n_1(\vec x') \otimes \frac{\vec x'-\vec x_2}{|\vec x'-\vec x_2 |^2} \, \D s.
$$
The above integral is evaluated in polar coordinates with pole \(\vec x_1\) and polar axis \(\vec e_1\):
$$
\vec x' = \vec x_1 + a_1 \bigl(\cos\varphi \, \vec e_1 + \sin\varphi \, \vec e_2\bigr),
$$
with \(s = a_1 \, \varphi\) and \(\vec n_1(\vec x') = \cos\varphi \, \vec e_1 + \sin\varphi \, \vec e_2\). Then
$$
\vec x' - \vec x_2 = a_1 \bigl(\cos\varphi \, \vec e_1 +\sin\varphi \, \vec e_2\bigr) - r_{12} \, \vec e_1
$$
and, upon integration, we obtain 
\begin{equation} \label{eq:utile_pour_trace}
  \begin{aligned}[b]
    \tens{\Gamma}_{12}^\infty
    ={}&  \frac{1}{2\pi\sigma_0} \, \frac{a_2^2}{a_1} \, \int_0^{2\pi} \bigl[ \cos\varphi \, \vec e_1 + \sin\varphi \, \vec e_2 \bigr] \\
    &\qquad\otimes \frac{a_1 \bigl(\cos\varphi \, \vec e_1 +\sin\varphi \, \vec e_2\bigr) - r_{12} \, \vec e_1}{\bigl\lvert a_1 \bigl(\cos\varphi \, \vec e_1 +\sin\varphi \, \vec e_2\bigr) - r_{12} \, \vec e_1 \bigr\rvert^2} \, \D \varphi\\
    ={}& \frac{1}{2\sigma_0} \, \frac{a_2^2}{r_{12}^2} \, \bigl(\vec e_2 \otimes \vec e_2 - \vec e_1 \otimes \vec e_1\bigr),
  \end{aligned}
\end{equation}
which delivers the intrinsic expression~\eqref{eq:20201007151853}.

\subsection{Evaluation of the self-influence tensor \(\tens{\Gamma}_{22}^\infty\)}

Inside the inclusion \(\Omega_2\), we infer from~\eqref{eq:20201006160008} that the electric field is constant and reads $\displaystyle \vec E = -\frac{\vec P}{2\sigma_0}$, which implies that
$$
\tens{\Gamma}_{22}^\infty = \frac{\tens{I}}{2\sigma_0}.
$$

\section{Numerical validation of formula~\eqref{eq:20220719170815}} \label{sec:20220719172106}

Note: the following numerical validation can be reproduced from the dataset  at the \texttt{Recherche Data Gouv} open data repository~\citep{bris2023a}.

\medskip

\begin{figure}
  \centering
  \includegraphics[width=0.5\linewidth, bb= 17.9cm 4.4cm 41cm 27.6cm]{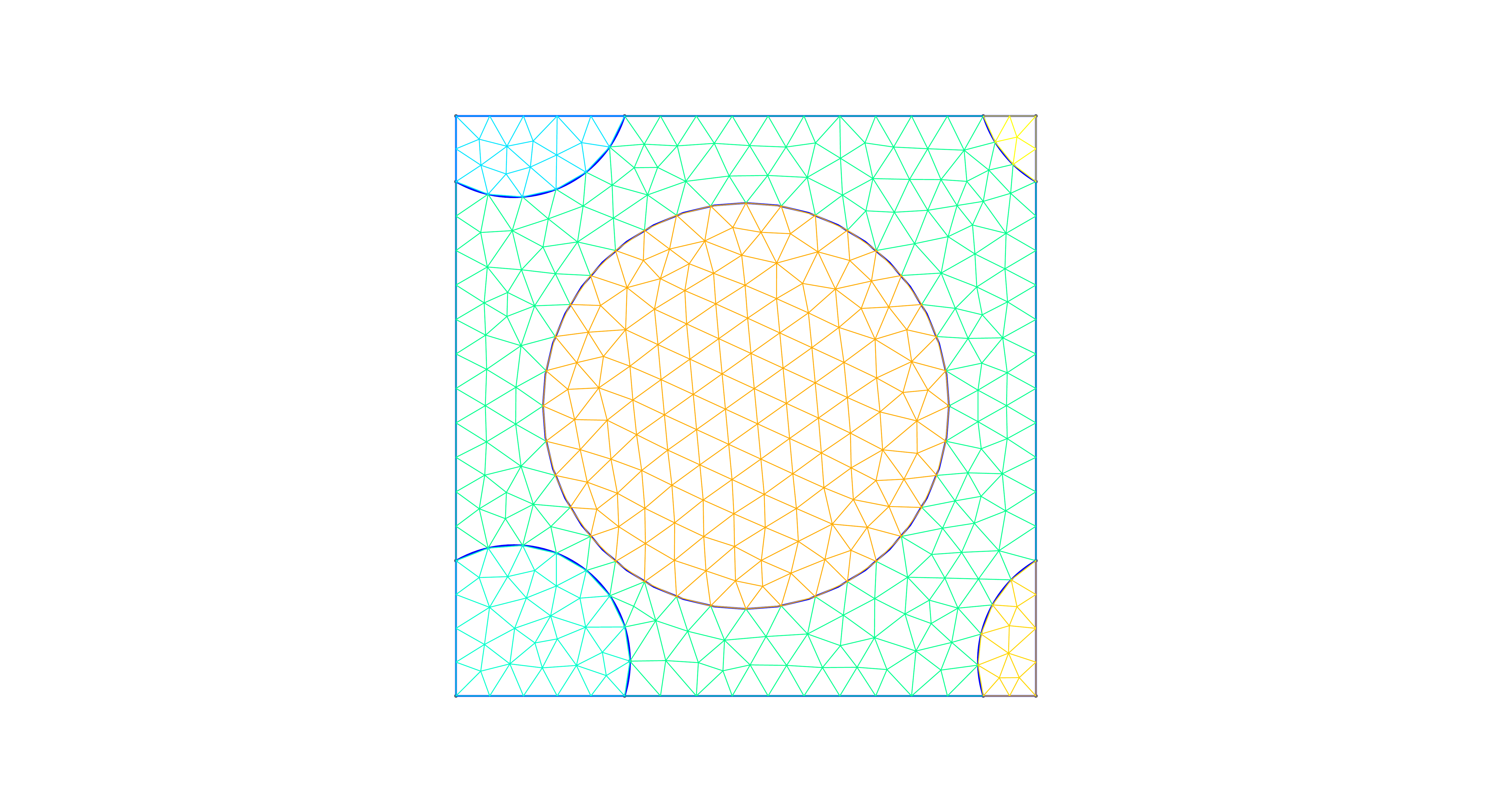}
  \caption{The configuration considered for the numerical validation of~\eqref{eq:20220719170815} (see Appendix~\ref{sec:20220719172106}). The mesh-size is \(h=L/16\) in the case shown on the figure. \label{fig:20220721101425}}
\end{figure}

In the relation~\eqref{eq:20220719170815}, we have proposed a correction to the expression~\eqref{eq:20220707152227} initially introduced by~\citet{zece2021}. The validity of this correction has not been proved mathematically. However, we provide in this appendix a numerical validation of this formula for the set of parameters displayed in Table~\ref{tab:20220721120422}. The resulting configuration is represented in Fig.~\ref{fig:20220721101425}.

\begin{table}
  \centering
  \begin{tabular}{c|c|l}
    Symbol & Value & Comment\\
    \hline
    \(L\) & 1.0 & Size of the square unit-cell\\
    \(\sigma_0\) & 1.0 & Uniform conductivity\\
    \(\vec x_1\) & \((0.5L, 0.5L)\) & Center of disk 1\\
    \(a_1\) & \(0.35L\) & Radius of disk 1\\
    \(\vec x_2\) & \((0.1L, 0.06L)\) & Center of disk 2\\
    \(a_2\) & \(0.2L\) & Radius of disk 2\\
    \hline
  \end{tabular}
  \caption{Parameters used for the numerical validation of~\eqref{eq:20220719170815} (see Appendix~\ref{sec:20220719172106}). \label{tab:20220721120422}}
\end{table}

\smallskip

The left-hand side of~\eqref{eq:20220719170815} (``true'' periodic influence tensor) is evaluated by means of a finite element simulation (see Fig.~\ref{fig:20220721101425}) with increasing element fineness (from \(L/4\) to \(L/512\)).

\begin{figure}
  \centering
  \includegraphics{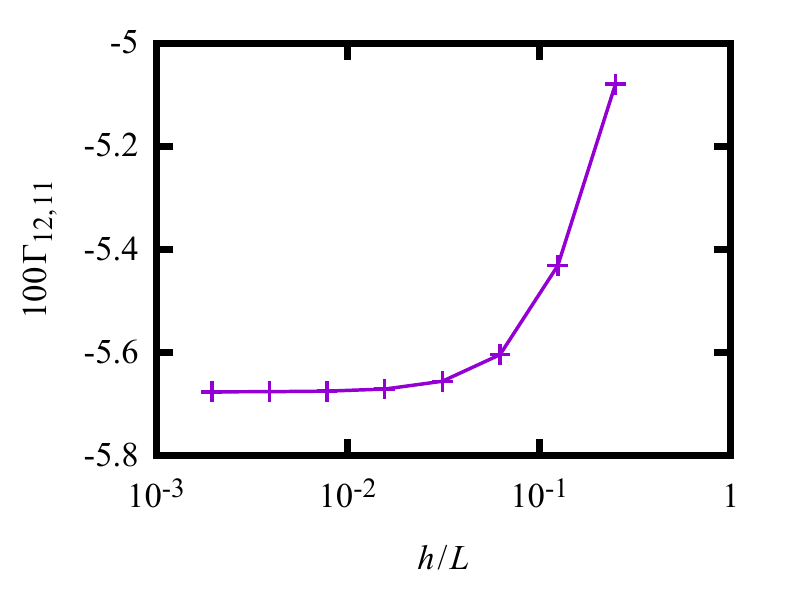}
  \caption{FEM approximation of the \((1,1)\) coefficient of the influence tensor \(\tens{\Gamma}_{12}\), as a function of the mesh size \(h\). To improve visibility, we plot $100 \, \Gamma_{12, 11}$ rather than $\Gamma_{12, 11}$. \label{fig:20220720121116}}
\end{figure}

\begin{figure}
  \centering
  \includegraphics{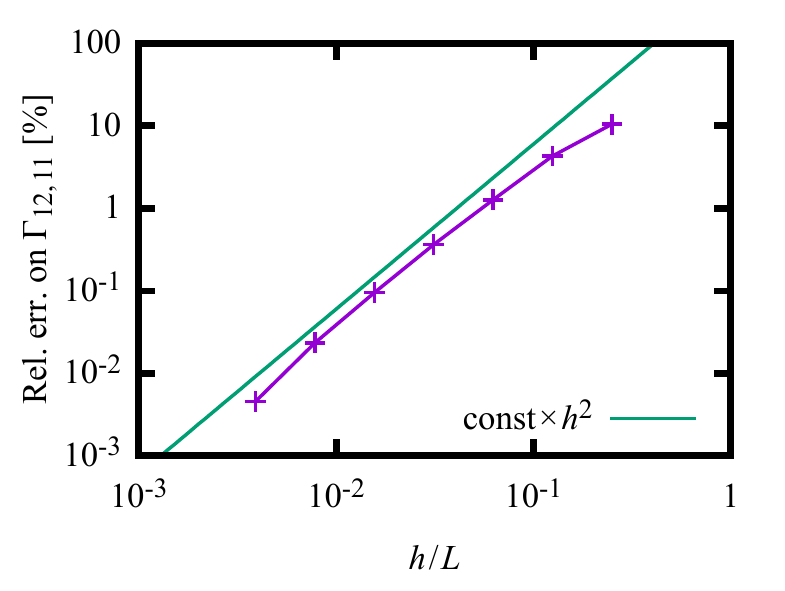}
  \caption{Relative error on the \((1,1)\) coefficient of the influence tensor \(\tens{\Gamma}_{12}\) (when computed by FEM), as a function of the mesh size \(h\). The straight line represents the convergence rate \(h^2\). \label{fig:20220720121132}}
\end{figure}

The variational problems to be solved are: find \(\Phi_i\), $i=1,2$, such that, for all \(\Psi\), 
$$
\int_\cell \grad\Phi_i \cdot \grad\Psi = \int_{\Omega_2 \cap \cell} \partial_i \Psi,
$$
where \(\Phi_1\), \(\Phi_2\) and \(\Psi\) are $\cell$-periodic. The coefficients of \(\tens{\Gamma}_{12}\) are then computed as
$$
\Gamma_{12, ji} = \vec e_j \cdot \tens{\Gamma}_{12} \, \vec e_i = \frac{1}{\sigma_0 \, | \Omega_1 |} \int_{\Omega_1} \partial_j \Phi_i.
$$
The values of \(\Gamma_{12,11}\) are plotted in Fig.~\ref{fig:20220720121116} versus the element-size \(h\). Convergence as \(h \to 0\) seems to be observed. This is confirmed by Fig.~\ref{fig:20220720121132}, which represents the relative error on \(\Gamma_{12, 11}\), defined by
$$
\left| \frac{\Gamma_{12, 11}(h) - \Gamma_{12, 11}(h_{\min})}{\Gamma_{12, 11}(h_{\min})} \right|,
$$
where the reference value is taken as the FE result for the finest mesh size considered here, that is \(h_{\min}=L/512\). The figure clearly shows the expected \(h^2\) convergence rate. 

\medskip

\begin{figure}
  \centering
  \includegraphics{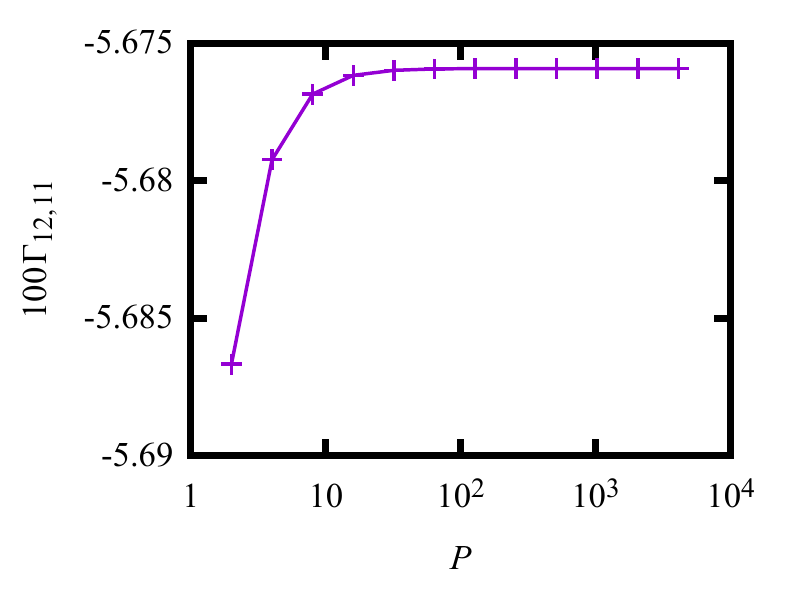}
  \caption{Estimates of the \((1,1)\) coefficient of the influence tensor \(\tens{\Gamma}_{12}\) based on the Poisson summation formula, as a function of the truncation parameter \(P\). As in Fig.~\ref{fig:20220720121116}, to improve visibility, we plot $100 \, \Gamma_{12, 11}$ rather than $\Gamma_{12, 11}$. \label{fig:20220720122123}}
\end{figure}

\begin{figure}
  \centering
  \includegraphics{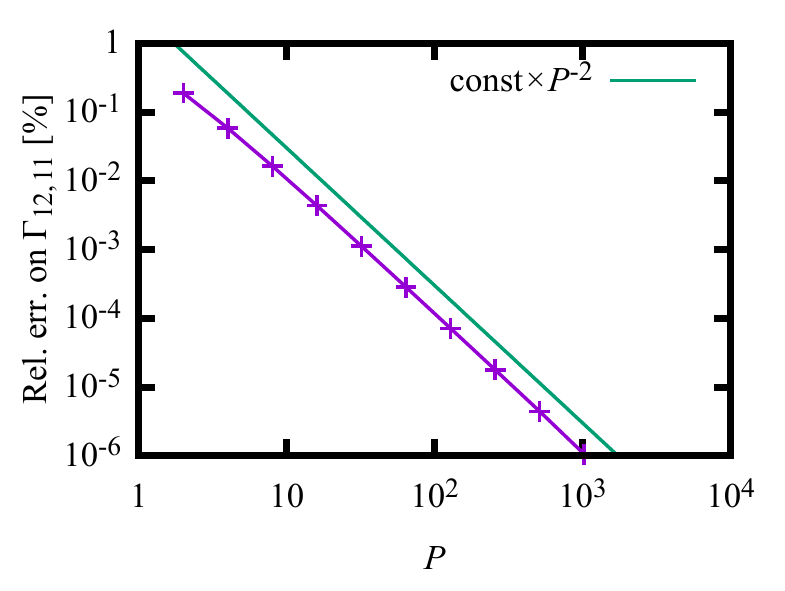}
  \caption{Relative error on the \((1,1)\) coefficient of the influence tensor \(\tens{\Gamma}_{12}\) (when computed by the Poisson summation formula), as a function of the truncation parameter \(P\). The straight line represents the convergence rate \(P^{-2}\). \label{fig:20220720122205}}
\end{figure}

We now turn to the right-hand side of~\eqref{eq:20220719170815} (Poisson summation). This right-hand side is estimated for various values of $P$ (ranging from 2 to 4096), using~\eqref{eq:20201007151853}. The obtained values of the \((1,1)\) coefficient of the influence tensor are plotted in Fig.~\ref{fig:20220720122123}. Again, convergence seems to be observed as \(P \to +\infty\). This is confirmed by Fig.~\ref{fig:20220720122205}, which represents the relative error on \(\Gamma_{12, 11}\), now defined by 
$$
\left| \frac{\Gamma_{12, 11}(P) - \Gamma_{12, 11}(P_{\max})}{\Gamma_{12, 11}(P_{\max})} \right|,
$$
where the reference value is taken as the truncated sum for the largest value of \(P\) considered here, that is \(P_{\max}=4096\). The truncated sum seems to converge as \(P^{-2}\).

\medskip

The FEM (resp. Poisson) value of the complete $2 \times 2$ matrix \(\tens{\Gamma}_{12}\) for the smallest grid-size \(h\) (resp. for the largest truncation parameter \(P\)) is
\begin{align*}
  \tens{\Gamma}_{12} &= \begin{bmatrix}
                        -0.0567582 & -0.0113464\\
                        -0.0113464 & -0.0689035
                      \end{bmatrix} && \text{(FEM),}\\
  \tens{\Gamma}_{12} &= \begin{bmatrix}
                        -0.0567591 & -0.0113466\\
                        -0.0113466 & -0.0689045
                      \end{bmatrix} && \text{(Poisson).}
\end{align*}
Both results are in excellent agreement, which provides a first validation of the proposed corrected formula~\eqref{eq:20220719170815}.

\section*{Acknowledgements}

This work has benefited from the French government grant ANR-11-LABX-022-01 (Labex MMCD, Multi-Scale Modelling \& Experimentation of Materials for Sustainable Construction) managed by ANR within the frame of the national program Investments for the Future.




\section*{Software}

The meshes required by the full-field simulations have been generated with the \texttt{Gmsh} software~\citep{geuz2009}, using constructive geometry primitives provided by the \texttt{Open\-Cascade} geometry kernel\footnote{\url{https://dev.opencascade.org/}, last retrieved 2022-07-27.}. The \texttt{FreeFem++} software~\citep{hech2012} has been used for all FEM simulations. All other computations have been performed with the Julia programming language~\citep{beza2017}.

\bibliography{EIMVarRed}

\end{document}